\documentclass[aps,pre,preprint,groupedaddress,showkeys]{revtex4-1}

\usepackage{hyperref}
\usepackage{graphicx}
\usepackage{epsfig}
\usepackage{bm}
\usepackage{verbatim}
\usepackage{amssymb,amsfonts,amsmath}

\begin{document}

\title{ Hydrophobic force a Casimir-like effect due to \\ hydrogen bond fluctuations }

\author{Jampa Maruthi Pradeep Kanth}

\email[]{jmpkanth@imsc.res.in}
\affiliation{The Institute of Mathematical Sciences, C.I.T.Campus, Tharamani, Chennai  600113,  India.}

\author{Ramesh Anishetty}
\email[]{ramesha@imsc.res.in}
\affiliation{The Institute of Mathematical Sciences, C.I.T.Campus, Tharamani, Chennai  600113,  India.}

\date{\today}

\begin{abstract}

Hydrophobic force, interfacial tension, transverse density profile in confined water system are addressed from first principles of statistical mechanics in a lattice model for water. Using molecular mean field theory technique we deduce explicit expressions for each of the above mentioned phenomena and show that hydrophobic force is a manifestation of Casimir-like effect due to hydrogen bond fluctuations in confined water. It is largely influenced by the long range correlations of orientational fluctuations. All the computations are parameter-free and we compare favorably with results of molecular dynamics simulations and experiments.

\end{abstract}



\maketitle

\section{Introduction} {\label{section_introduction}}

Hydrophobic force causes attraction between non-polar solutes in liquid water and is known to be an important driving force in micellar aggregation, cell membrane formation, large protein structures \citep{BallChemRev2008,*PrivalovAdvProtChem1988,*Tanford1997}. Although hydrophobic force is widely known to be pervading distinct physical, chemical and biological phenomena, its occurence as a solvent induced interaction was first suggested only in 1945, by Frank and Evans through calorimetric studies on hydrocarbons \citep{Frank1945} and later elucidated in biological context by Kauzmann \citep{Kauzmann1959}. The former study noted that transferring a small hydrophobe into water was accompanied by unfavorable free energy change \citep{Butler1937}, dominated by entropy reduction due to reorganization of vicinal water molecules \citep{Frank1945}. Hence, two hydrophobes show tendency to coalesce in order to minimize the unfavorable free energy. In the following years the applicability of this viewpoint for different sizes of hydrophobes such as alkanes, proteins has been discussed \citep{KlotzScience1958,*ScheregaJCP1962,*Engberts1993}. Theoretical and simulational investigations were limited to small solutes and interaction between them on scales of few Angstroms \citep{PrattJCP1977,*BerneJCP1979,*StillingerJCP1979,*TanakaJCP1987,*Herman1993}. In early 1980s the first direct measurement of an attractive force between hydrophobic surfaces has been carried out using surface force apparatus \citep{IsrNature1982}. Large hydrophobic surfaces made up of chemisorbed mica were employed in a crossed cylinder geometry and the measured force between them was related to interaction free energy per unit surface area using Derjaguin approximation \citep{Isrbook1992}. The force was seen to be influential upto hundreds of Angstroms and stronger than inter-surface van der Waal forces \citep{IsrNature1982}. The qualitative aspect of the result i.e., long range and monotonic nature of the force withstood the test of time \citep{Claesson2001,MeyerPNAS2006,Hammer2010}. To understand the same, various physical mechanisms were proposed, some generic to any fluid such as metastability of confined fluid \citep{ClaessonScience1988,*LuzarPRL2003}, dewetting-induced cavitation under liquid-vapor coexistence conditions \citep{LumJPC1999}, fluid structuring effects \citep{MarceljaCPL1976,*Eriksson1989,*Besseling1997} and some, dependent on surface details like correlated dipolar fluctuations \citep{AttardJPC1989,*PodgornikJCP1989,*Tsao1993}, charged bilayer patches \citep{MeyerPNAS2005}, nanobubbles \citep{AttardLangmuir1996,*TyrellPRL2001,*AttardPhysicaA2002}. The above phenomenological descriptions, however, are either envisaged in narrow range of fluid conditions or specifically depend on hydrophobization patterns on surfaces. Besides, they were unsuccessful in reproducing generic features of hydrophobic force seen in experiments \citep{Claesson2001}. We attempt to understand in a model study the nature of hydrophobic force by analyzing correlations of hydrogen bond fluctuations in liquid water and the effect of modification caused by presence of surfaces on these fluctuations. The analysis is carried out at generic fluid conditions within the model and applies to generic surfaces.

We discuss some essential aspects in modelling water and past attempts in this direction. Hydrogen bonding is an orientation-dependent attraction between two water molecules \citep{StillingerScience1980}. In order to analyze hydrogen bond fluctuations it is therefore essential to envisage both density and orientational degrees of freedom of each water molecule. There are models galore proposed and analyzed to reproduce anomalous thermodynamic properties of water \citep{BellJoP1972,*BlumJCP1985,*JaglaJCP1999,*TruskettJCP1999,*StanleyJSP2003,Guillot2002,*Vega2009}. However, theoretical attempts to envisage fluctuations in water are limited to Ornstein-Zernike-like phenomenological approaches, wherein integral equations only in terms of density correlation were heuristically proposed and are numerically solved using different closure approximations \citep{HansenSimpleLiquids,*Gray1984,BlumJCP1990,*RichardiJCP1999}. Wertheim's theory of associating fluids envisages similar density correlations to be solved in compliance with steric constraints imposed by formation of associated n-mers \citep{WertheimJStatPhy1984Vol35,*Wertheim2JStatPhy1984Vol35,*WertheimJStatPhy1986Vol42}. The success of these descriptions crucially depends on approximation schemes employed \citep{HansenSimpleLiquids,*Gray1984,KolafaMolPhys1989,*EvansJCP2003,*SciortinoJCP2008,*DillJCP2009}. Other approaches specific to molecular fluids such as RISM were seen to be less predictive in case of associating fluids \citep{ChandlerARPC1978}. Also, perturbation theories based on smallness of interaction strength found little success in reproducing liquid phase properties of water-like fluids \citep{AndersenJCP1973Vol59}.

Quantum mechanical calculations on water dimer in gas phase and diffraction study of ice forms provide sufficient evidence for specific nature of hydrogen bonding \citep{Eisenberg1969}, i.e., positively polarized hydrogen atom covalently bonded to an oxygen interacts only with negatively polarized lone-pair of neighboring oxygen. The specificity necessitates the density of hydrogen bonds and dangling bonds (hydrogens or lone-pairs which are not hydrogen bonded) to be commensurate with molecular density, stated as a \textit{sum rule} \citep{KanthPhysica2011}. Consequently, density and orientational fluctuations (the latter being inherently connected to bond fluctuations) are not totally independent; their long wavelength fluctuations, especially, are to be consistent with the sum rule. Effective interaction models for water designed for numerical simulations (molecular dynamics) \citep{RahmanJCP1971Vol55,*StillingerJCP1974Vol60,Guillot2002,*Vega2009} provide successful instances of (implictly) envisaging most essential features of hydrogen bond fluctuations consistent with the sum rule. A water molecule is often modelled as a polar molecule with charges corresponding to hydrogens and lone-pairs placed at vertices of a tetrahedron \citep{RahmanJCP1971Vol55,*StillingerJCP1974Vol60,MahoneyJCP2000Vol112}. A complete description of molecular correlations in such models can be achieved by defining a set of orthonormal vectors in terms of atomic coordinates and defining correlations among them. Large scale molecular dynamics (MD) simulations at ambient conditions reveal that density correlations are short ranged and extend no further than $12$ \AA{}; whereas, dipolar orientations of water molecule, which are receptive to bond fluctuations in the neighborhood, are correlated over large distances, atleast upto $75$ \AA{} \citep{KanthPRE2010}. Two correlation lengths of order $5.2$ \AA{} and $24$ \AA{} were inferred. Coulomb interactions, surprisingly, have little effect on asymptotic behavior of these correlations \citep{KanthPRE2010}. It is also suggested that mesoscopic hydrophobic solutes when coupled to dipolar fluctuations in water medium experience a long range exponential force, which is attractive in nature and dependent on shape and mutual orientation of solute surfaces \citep{KanthPRE2010}. The exponential decay bears a striking consistency with that seen in experiments measuring hydrophobic force between large surfaces \citep{IsrNature1982,Claesson2001}. For the case of large hydrophobic surfaces, correlations in confined water need to be ascertained. To simulate such a system the surfaces need to be several times larger than the longest correlation length in the system in order to obtain proper equilibrium conditions. This requires huge system size that would render the simulation prohibitively resource intensive. In addition, the accompanying free energy change could be very small due to weak nature of correlations at large distances. Instead, we take analytic route to describe hydrogen bond fluctuations in water and the effect of spatial confinement on them.

Hydrophobes are known to interact unfavorably with water molecules in contrast to strong and cohesive water-water interactions. In addition, large surfaces substantially disrupt the hydrogen bond network whose fluctuations are suppressed at surface boundaries. The setting is ideally suited for fluctuation-induced force between the surfaces driven by thermal energy in confined water. Forces of this nature are generically called \textit{Casimir forces} as they were first discussed by Casimir in the case of electromagnetic fluctuations confined between conducting plates \citep{Casimir1948} which was later studied in detail by Lifshitz \citep{Lifshitz1956}. Such forces are now envisaged in widely different contexts \citep{KardarRMP1999}. Fisher and de Gennes argued that when a binary liquid mixture is confined between surfaces which have specific affinity towards one of the fluid components, Casimir-like density fluctuations in the liquid give rise to an effective force between surfaces \citep{Fisher1978}. Origin of the force is entropic in nature; in that, the free energy is increased due to restriction imposed on fluctuations by the boundaries, thereby system tends to minimize the separation in order to reduce the free energy cost. We investigate hydrophobic force to be a manifestation of Casimir-like force, here, due to density and orientational fluctuations in liquid water.

When water is confined between hydrophobic surfaces the inherent field fluctuations vanish on surfaces. Furthermore, water molecules at interface with each surface have restricted orientational entropy owing to repulsive hydrophobe-water interactions. This effect gives rise to modified fluctuations at each interface. We study the collective consequences of these effects on the overall free energy of the system in a model study. We define a simple water model wherein density and orientations of a water molecule are envisaged. The specific nature of hydrogen bonding between molecules is incorporated and associated constraints on the bond network are taken care of in the analysis. Partition function is solved about a mean field which is consistently deduced at arbitrary densities within the model. Fluctuation properties are also deduced. Large correlation length is seen for orientational fluctuations. Two macroscopic surfaces are envisaged as boundaries in a spatial dimension. The change in free energy in the system due to the presence of surfaces is calculated and is seen to be composed of three important contributions : (i) Casimir part, which arises solely from discretization of fluctuation modes between boundaries and is generic to all surfaces; (ii) Interfacial tension, which is free energy change due to modified fluctuations at hydrophobe-water interface. It is dependent on nature of surface-water interaction and to a small extent, also on separation distance between the surfaces; (iii) Interfacial fluctuations-induced free energy, which is due to correlations of modified fluctuations at both interfaces. It depends on type of both surfaces and their interaction with water. The results are discussed for different types of surfaces such as hydrophobic and hydrophilic. We find that the Casimir part is leading contribution and is an inverse power-law function of separation distance. However, numerically the magnitude of Casimir part is significant only upto four times the longest correlation length in the model. Interfacial tension also varies with separation distance, but its variation is numerically insignificant. Interfacial fluctuations-induced contribution is seen to be exponentially decaying with distance, analogous to the force form deduced for mesoscopic surfaces \citep{KanthPRE2010}. Furthermore, we find that all the contributions are of comparable order of magnitude consistent with experimental values. The dependence of the force on fluid conditions like temperature, average hydrogen bonds is also discussed. Our results indicate that hydrophobic force qualitatively imitates Casimir-like force behavior within our model study. It is desirable to emulate the computation within more realistic models of water possibly with the help of MD simulations. We also looked at transverse density profile for confined water and show that an increase in density occurs near interfaces.

\section{Model for water} {\label{section_watermodel}}

We envisage our water model in the background of a lattice to exactly account for hard-sphere repulsion between atoms at short distances. Also, constraints of hydrogen bond network are explicitly taken care of in lattice background. We developed molecular mean field (MMF) technique in an earlier paper \citep{KanthPhysica2011} to address the same model in the infinite volume limit. We present here a brief summary of the model and MMF technique and then, address the case of confined water.

On a three dimensional hypercubic lattice we define occupation field $W(r) = \lbrace 0, 1 \rbrace$ corresponding to water being absent or present, respectively, at the site $r = (x,y,z)$. At each \textit{occupied} site we define bond arm field $H_{\alpha}(r)$ which resides on the links around the site $r$. $H_{\alpha}(r) = \lbrace 0, \pm 1 \rbrace$, where $\alpha = \lbrace \pm 1, \pm 2, \pm 3 \rbrace$ correspond to six directions around the site. $H_{\alpha}(r) = 1$  refers to hydrogen arm on the corresponding link, $-1$ to lone pair arm and $0$ for no arm. The constraints between $W(r)$ and $H_{\alpha}(r)$ being,
	\begin{align}
		\sum_{\alpha} H_{\alpha}^{2}(r) \ &= \ 4 W(r) \label{H2constraint}
		\\ \sum_{\alpha} H_{\alpha}(r) \ &= \ 0 \label{Hconstraint}
	\end{align}
which imply that every water molecule has two hydrogen and two lone-pair arms only. A hydrogen bond is realized when two water molecules two lattice units apart have one of each's hydrogen and lone-pair arms meet at a site, as shown in Fig.(\ref{fig_allowedconfigs}). When two molecules are on near-neighbor sites they are disallowed to have any non-zero bond arm on the link between them. The constraint is given by :
	\begin{equation}
		W(r) \left( \sum_{\alpha} H^{2}_{\alpha}(r + e_{\alpha}) \right) = 0 \label{unoccupied_Hconstraint}
	\end{equation}
We write a general interaction Hamiltonian in terms of $H_{\alpha}$ field as below :
	\begin{equation}
		{\cal H} \ = \ \frac{\tilde{\lambda}}{2} \sum_{r} \sum_{\alpha, \alpha^{'}} H_{\alpha}(r - e_{\alpha}) H_{\alpha^{'}}(r - e_{\alpha^{'}})
	\end{equation}
where, $\tilde{\lambda}$ is interaction strength and $\alpha$, $\alpha^{'}$ denote directions around site $r$. There are additional restrictions on $H_{\alpha}$ field, namely,
\\ (i) at any site no more than two bond arms meet i.e.,
	\begin{equation}
		0 \leq \sum_{\alpha} H^{2}_{\alpha}(r + e_{\alpha})  \leq  2 \label{unoccupied_H2values}
	\end{equation}	 		
(ii) two non-zero bond arms are disallowed from meeting at a site i.e., anti-bonds are disallowed, 
 	\begin{equation}
		-1 \leq \sum_{\alpha} H_{\alpha}(r + e_{\alpha})  \leq 1 \label{unoccupied_Hvalues}
	\end{equation}	 		
The grand canonical partition function for the system at a finite chemical potential $\tilde{\mu}$ for water and inverse temperature $\beta$ is given by :
	\begin{equation}
		Z  = \prod_{r} \sum_{\displaystyle W(r), H_{\alpha}(r)}^{'} \exp \left[ - \beta \sum_{r} \left( {\cal H} - \tilde{\mu} W(r) \right) \right] \label{partitionfunction}
	\end{equation}
where, the prime over summation indicates that the $W(r)$ and $H_{\alpha}(r)$ sum have to be carried out in compliance with  Eqs.(\ref{H2constraint},\ref{Hconstraint},\ref{unoccupied_Hconstraint},\ref{unoccupied_H2values},\ref{unoccupied_Hvalues}). Evaluating $Z$ amounts to enumerating all possible bond configurations that satisfy above constraints and calculating the exponential in Eq.(\ref{partitionfunction}) for those configurations over allowed range of $W$ and $H_{\alpha}$ at fixed values of $\tilde{\mu}$, $\beta$ and $V$ the volume of the system.

\begin{figure*}[h]
		\epsfig{file= 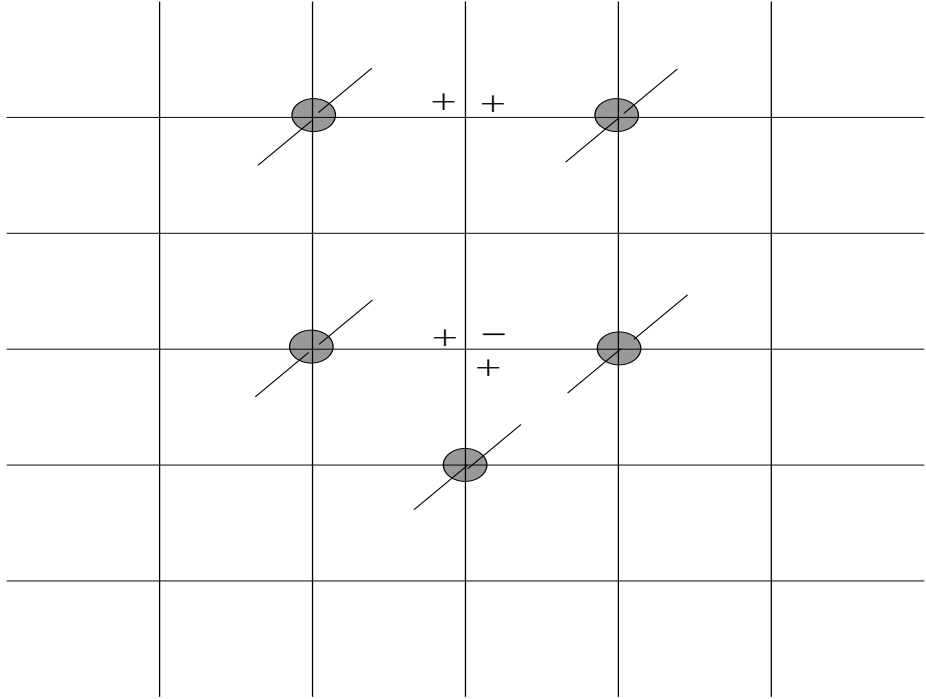,width=0.75\linewidth,height=0.4\textheight}
		\caption{\label{fig_disallowedconfigs} Disallowed configurations : non-zero bond arms of same type of two molecules meeting at a site; more than two non-zero arms meeting at a site.}
	\end{figure*}

The restrictions represented by Eqs.(\ref{unoccupied_Hconstraint},\ref{unoccupied_H2values},\ref{unoccupied_Hvalues}) are at sites where there is no water. These are shown in Fig.(\ref{fig_disallowedconfigs}). To implement them in our analysis it is useful to define two discrete integer fields $b(r)$, $q(r)$ :
	\begin{subequations}
	\label{bqdefinition}
	\begin{align}
		b(r) \ & = \ \sum_{\alpha} H_{\alpha}^{2}(r+{e}_{\alpha})
		\\ q(r) \ & = \ \sum_{\alpha} H_{\alpha}(r+{e}_{\alpha})
	\end{align}
	\end{subequations}
The discrete field $b(r)$ counts the number of non-zero arms in the neighborhood of site $r$, while $q(r)$ measures the charge i.e, difference between number of hydrogen arms and lone-pair arms meeting at site $r$. By construction, $b(r)$ varies between $0$ and $6$  on a three dimensional hyper-cubic lattice and $q(r)$ in turn varies between $-b(r)$ to $b(r)$. By imposing the condition that $b(r) \leq 2$ in our analysis we ensured that no more than two arms can meet at a site. Furthermore, for $b(r) = 2$ we demand $q(r) = 0$ to disallow anti-bond configurations. In terms of these variables, Eqs.(\ref{unoccupied_Hconstraint},\ref{unoccupied_H2values},\ref{unoccupied_Hvalues}) can be rewritten as :
	\begin{align}
		  W(r) b(r)  \ & = \   0 	\label{bconstraint}
		 \\ (b(r), q(r)) \ & = \ \left\lbrace (0,0), (1,1), (1,-1), (2,0) \right\rbrace \label{bqvalues}
	\end{align}
where, $b(r)$ and $q(r)$ values are restricted only to the above set of mutually exclusive pairs. We now rewrite the partition function as below :
	\begin{equation}
		Z \ = \ \prod_{r} \sum_{\displaystyle  { W(r), H_{\alpha}(r)} \atop \displaystyle { b(r), q(r) } }^{'} \exp \left[ - \beta \sum_{r} \left( {\cal H} - \tilde{\nu} q^{2}(r) - \tilde{\mu} W(r) \right) \right]
	\end{equation}
where, we have additionally introduced a chemical potential $\tilde{\nu}$ for dangling bond configuration i.e., $(b,q) = (1, \pm 1)$. The fields $b$, $q$ are summed over allowed range given in Eq.(\ref{bqvalues}) and the prime over summation indicates that Eqs.(\ref{H2constraint},\ref{Hconstraint},\ref{bqdefinition},\ref{bconstraint}) act as constraints in the evaluation. Note that, since only hydrogen bond interaction is envisaged in the model, the Hamiltonian ${\cal H}$ can be rewritten as a simple expression :
	\begin{equation}
		{\cal H} \ = \  - \tilde{\lambda} \sum_{r} \ \delta({b(r),2}) \label{Hamiltonian_redefinition}
	\end{equation}
where, Kronecker delta function denoted here as $\delta({p,q})$ is defined as $\delta({p,q}) = 1$ for $p = q$ and $0$ otherwise. All the possible hydrogen bond configurations are implied from solving the non-local constraints Eq.(\ref{bqdefinition}). These constraints are enforced in the partition function by introducing auxiliary fields, as given below :
	\begin{subequations}
	\label{auxiliaryfields}	
	\begin{align}
		 & \delta\left( b(r), \sum_{\alpha} H^{2}_{\alpha}(r + {e}_{\alpha}) \right) = \frac{1}{2N+1} \sum_{\displaystyle \eta(r)} \exp\left[ -i \frac{\pi}{N} \eta(r) \left( b(r) - \sum_{\alpha} H^{2}_{\alpha}(r + {e}_{\alpha}) \right) \right]
		 \\ & \delta\left( q(r), \sum_{\alpha} H_{\alpha}(r + {e}_{\alpha}) \right) = \frac{1}{2N+1} \sum_{\displaystyle \phi(r)} \exp\left[ - i \frac{\pi}{N} \phi(r) \left( q(r) - \sum_{\alpha} H_{\alpha}(r + {e}_{\alpha}) \right) \right]
	\end{align}
	\end{subequations}
where, $\eta(r)$ and $\phi(r)$ act as dual fields to density and charge of bond arms in a local neighborhood. The discrete $\eta$ and $\phi$ fields take integer values in the range $[-N, N]$ at every site, where $N$ is any suitably large integer (greater than $8$).

The partition function can be rewritten in terms of new variables and auxiliary fields as :
	\begin{align}
		\nonumber & Z \ =  \ \left[ \prod_{r}  \sum^{'}_{\displaystyle W(r), H_{\alpha}(r) \atop \displaystyle b(r), q(r)} \frac{1}{(2N+1)^{2}} \sum_{\displaystyle \eta(r), \phi(r)} \right] \exp  \sum_{r} \left[ \vphantom{\sum_{\alpha}} -\beta ( {\cal H}  -  \tilde{\nu} q^{2}(r) - \tilde{\mu} W(r) ) \right. 
		\\ &  + \left. {i \frac{\pi}{N}} {\displaystyle \eta(r) \left( \sum_{\alpha} H^{2}_{\alpha}(r + {e}_{\alpha}) - b(r) \right) }  + {i \frac{\pi}{N}} \phi(r) \left( \sum_{\alpha} H_{\alpha}(r + {e}_{\alpha}) - q(r) \right) \right] \label{Zequation}
	\end{align}
Here, prime over summation refers to sum being restricted to local on-site constraints Eqs.(\ref{H2constraint},\ref{Hconstraint},\ref{bconstraint}) only. The introduction of auxiliary fields $\eta(r)$ and $\phi(r)$ allows summation over other discrete fields $W(r), H_{\alpha}(r), b(r), q(r)$ within their respective allowed range at each site \textit{without} any restriction from the neighborhood configurations i.e., as if a single site functional $Z_{site}$, as given below : 
	\begin{equation}
		Z  \ = \   \int [{\cal D} \eta] [{\cal D}\phi] \prod_{r} Z_{site}(\eta(r), \phi(r), \nabla_{\alpha}\eta, \nabla_{\alpha}\phi) \label{Zequation_v2}
	\end{equation}
where, the summation over $\eta, \phi$ fields is transformed into an integral in the limit of $N \rightarrow \infty$ and $Z_{site}$ is given as sum of weights corresponding to each allowed state i.e., void, dangling bond, hydrogen bond and water. It is given by,
	\begin{align}		
		  Z_{site} \ & = \ 1 + 2 \nu e^{-i \eta(r)} \cos(\phi(r)) + \lambda e^{-2 i \eta(r)} + \mu   C(\eta, \phi, \nabla_{\alpha}\eta, \nabla_{\alpha}\phi) )  \label{zsite}
		\\  C(\eta, \phi, \nabla_{\alpha}\eta, \nabla_{\alpha}\phi) & =  \sum^{'}_{\displaystyle \stackrel{\displaystyle H_{\alpha} = 0,\pm 1}{\alpha = \pm 1, \pm 2, \pm 3}} \exp\left[ i \sum_{\alpha} ( H^{2}_{\alpha}(r) {\eta}(r+e_{\alpha}) + H_{\alpha}(r) {\phi}(r+e_{\alpha}) )  \right]  \label{Cweight}
	\end{align}
where, $\nu \equiv \exp(\beta \tilde{\nu})$, $\lambda \equiv \exp(\beta \tilde{\lambda})$ and $\mu \equiv \exp(\beta \tilde{\mu})$ are fugacities of dangling bond, hydrogen bond and water states, respectively. The orientational degrees of freedom of water yields $C(\eta, \phi, \nabla_{\alpha}\eta, \nabla_{\alpha}\phi)$ given by Eq.(\ref{Cweight}), where the summation is over orientations at site $r$. The prime over summation indicates $H_{\alpha}$'s of each orientation satisfy Eqs.(\ref{H2constraint},\ref{Hconstraint}). The exponential corresponds to an orientation and it is a function of dual fields at near-neighbor sites towards which non-zero bond arms of the orientation are directed. The densities of dangling bond (DB), hydrogen bond (HB) and water ($\rho$) are calculated from partial derivative of partition function with respect to $\beta \tilde{\nu}$, $\beta \tilde{\lambda}$, $\beta \tilde{\mu}$, respectively.

\subsection{MMF theory} {\label{section_MMF}}

The partition function has a unique maximum at isotropic and homogeneous field configuration $\eta = \phi = 0$. $Z_{site}$ at the maximum is given by $Z_{o}$ :
	\begin{equation}
		Z_{o} = (1 + 2 \nu + \lambda + 90 \mu)
	\end{equation}
The extremization condition $\sum_{x}({\delta}/{\delta \eta}) Z = 0$ implies the sum rule of the system i.e., $\textnormal{DB} + 2 \textnormal{HB} = 4 \rho$ exactly \citep{KanthPhysica2011}, while extremization with respect to $\phi$ field is trivially satisfied. To the zeroth order, partition function is $Z = (Z_{o})^{V}$ and sum rule translates as : 
	\begin{equation}
		2 \nu + 2 \lambda \ = \ 4(90 \mu)
	\end{equation}
Using this relation, the densities of dangling bond, hydrogen bond and water are given upto zeroth order as :
	\begin{subequations}
	\label{densities_zerothorder}
	\begin{align}
		\textnormal{DB} \ & \equiv \ \nu \frac{\partial }{\partial \nu}(\ln Z) \ = \ \frac{ 4\nu }{ 2 + 5\nu + 3 \lambda }
		\\ \textnormal{HB} \ & \equiv \  \lambda \frac{\partial }{\partial \lambda}(\ln Z) \ = \ \frac{ 2 \lambda }{ 2 + 5 \nu + 3\lambda }	
		\\ \rho \ & \equiv \ \mu \frac{\partial }{\partial \mu}(\ln Z) \ = \ \frac{ \nu + \lambda }{ 2 + 5 \nu + 3\lambda }
	\end{align}
	\end{subequations}
Eliminating $\lambda$ from equations for DB, HB we obtain :
	\begin{equation}
		\textnormal{HB} \ = \ 2 \rho - \frac{ \nu }{ \nu + 1 } (1 - 3 \rho)
		\label{equationofnetwork}
	\end{equation}
We call Eq.(\ref{equationofnetwork}) the \textit{equation of network}. It is a manifestation of sum rule in terms of model parameters. We choose dangling bond energy parameter to be zero i.e. $\nu = 1$ and measure temperature ($\beta^{-1}$) in units of hydrogen bond strength ($\tilde{\lambda}$). To zeroth order the theory is now parameter-free and all densities can be obtained as a function of temperature only. The equation of network can also be written in terms of average hydrogen bonds per molecule $h \equiv \frac{2 \textnormal{HB}}{\rho}$ as :
	\begin{equation}
		h = 7 - \frac{1}{\rho} \label{equationofnetwork_h_rho}
	\end{equation}
From zeroth order partition function the mean field free energy $G_{m}$ per unit volume can be given in terms of densities as :
	\begin{equation}
		\beta G_{m} = \ln( 1 - 5 \rho + HB )
		\label{equationofstate}
	\end{equation}
Eq.(\ref{equationofstate}) is analogous to equation of state for a system of hard spheres. It correctly predicts the density saturation in the model at $\rho = \frac{1}{3}$, $\textnormal{HB} = \frac{2}{3}$ or $h = 4$. Thus, equation of network is a manifestation and density saturation effect is a direct consequence of the sum rule.

Using Eq.(\ref{zsite}) for $Z_{site}$ we expand dual fields upto quadratic order about their maximum, perform Fourier transform on their fluctuations in a large cubic box using periodic boundary conditions, then integrate the resulting Gaussian functional in Eq.(\ref{Zequation_v2}) over all field configurations. This yields total free energy per unit volume $G$ upto one-loop correction which includes leading contributions due to fluctuations in density and orientations :
	\begin{equation}
		\beta G \ = \   \beta G_{m} +  \frac{1}{2} \int\limits^{\pi}_{-\pi} \frac{d^{3}k}{(2 \pi)^{3}} \ln \left(P_{\eta \eta}(\vec{k}) P_{\phi \phi}(\vec{k}) \right) \label{oneloop}
	\end{equation}
where,
	\begin{subequations}
	\label{propagators}
	\begin{align}
		P_{\eta \eta}(\vec{k}) & = \displaystyle 64 \rho \left(\frac{9}{10} - \rho \right) 
		\left[ \left(\Delta -  \frac{9}{20 (\frac{9}{10} - \rho)} \right)^{2} + \frac{3 (\frac{9}{25} - \rho)}{8 (\frac{9}{10} - \rho)^{2}} \right] 
		\\ P_{\phi \phi}(\vec{k}) & = \displaystyle \frac{96 \rho}{5} \left[ \Delta (1 -\Delta)  + \frac{5 \text{(DB)}}{96 \rho} \right] 
	\end{align}
	\end{subequations}
and $\Delta = \frac{1}{6} \sum_{i=1}^{3} (1 - \text{cos}(k_{i}))$; $k_{i}$ are vector components of $\vec{k}$. $P_{\eta \eta}$, $P_{\phi \phi}$ are fluctuation propagators of dual fields. The above expressions for propagators are simplified to the leading order using sum rule. Precise expressions in terms of original fugacities are given in Appendix (\ref{app_correlationfunctions}). The term involving propagators in the free energy expression (Eq.\ref{oneloop}) is the entropy contribution about the mean field. 

The correlation functions in the system in the momentum space are given by propagators $P_{\eta \eta}$ and $P_{\phi \phi}$. Density correlations are dominated by $\eta(r)$ correlations. They display coordination peaks in position space reminiscent of radial distribution function of fluids \citep{HansenSimpleLiquids} and do not have any long distance behavior (Fig.\ref{fig_h_beta}). Orientational correlations are dominated by $\phi(r)$ correlations at large distances. They display a correlation length of upto $4$ lattice units in liquid phase (Fig.\ref{fig_h_beta}).

We also envisaged Coulomb interaction between bond arm charges via a new dual field that couples to the charges. We find that Coulomb interactions have little effect on the asymptotic behavior of orientational fluctuations and also on MMF results like equation of network and equation of state \citep{KanthPhysica2011}.

\section{Water confined between macroscopic surfaces} {\label{section_confinedwater}}

We now study the case of water confined between two macroscopic hydrophobic surfaces. As a result of confinement the structure of fluctuations in the system is restricted by the boundaries, thereby causing entropy to be a function of separation distance between surfaces. In analogy with Casimir interaction, the distance-dependent entropy component of free energy of confined water leads to an effective interaction between hydrophobic surfaces. In addition, due to hydrophobe-water interactions, orientational fluctuations are modified at interface of each surface. The modified fluctuations and their correlations lead to interfacial tension proportional to area of surface and an induced interaction between the surfaces. The net effect is an interaction force that acts over distances longer than typical hydration structure of water. We utilize MMF framework to analyze these effects in a unified fashion within the proposed water model.

\begin{figure*}[h]
		\epsfig{file= 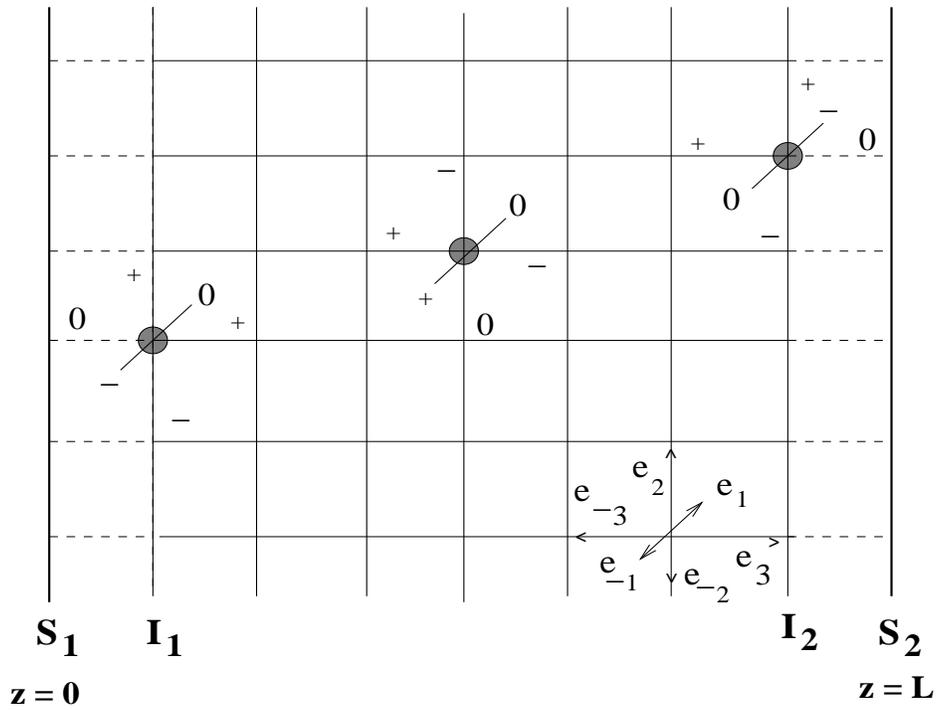,width=0.75\linewidth,height=0.4\textheight}
		\caption{\label{fig_confinedwater} Water confined between macroscopic surfaces. $S_{1}$, $S_{2}$ are surface planes at $z=0$ and $z=L$ respectively; $I_{1}$, $I_{2}$ are their respective interfaces at $z=1$ and $z = L-1$.}
	\end{figure*}

We envisage surfaces in the $(x,y)$ plane of rectangular coordinate system; one present at $z = 0$ and other at $z = L$ (Fig.\ref{fig_confinedwater}). Each surface excludes water from its region of occupation. Hence, $W = 0$ on surface sites. On the immediate layer, i.e., at $z = 1$ or $z = L-1$ called \textit{interface} layer,  water can be present and can take various orientations. For a hydrophobic surface if a non-zero bond arm of interface water is directed towards the surface, there would be a dangling bond on surface site; else a void state occurs. There can never be a hydrogen bond on surface i.e., $b \neq 2$ on surface. We will take care of these possibilities explicitly in our analysis. Consequently, we need not introduce $\eta$ and $\phi$ integrals (Eq.\ref{auxiliaryfields}) on the surface. Alternatively, we set $\eta = \phi = 0$ on surfaces.

The calculation of partition function begins with formulating the site functional $Z_{site}$ at each site, which comprises weights corresponding to each allowed state in the model. The site functional for all the sites in bulk region is of same form as given by Eq.(\ref{zsite}). On interface sites, weights corresponding to void state, dangling bond and hydrogen bond states remain unaltered. When a water molecule is present on a interface site its bond arms can orient in all possible ways. Only if one of the arms is towards the surface we assign a weight $\exp(\beta \tilde{\nu}_{S})$ to the corresponding orientation. For an ideal hydrophobic surface i.e., which is indifferent to bond arms of vicinal water, $\tilde{\nu}_{S} = 0$ (in general, $\tilde{\nu}_{S}$ can be positive or negative). Consequently, orientational weights for a water state on interface (with surface in $e_3$ direction) are given by :
	\begin{align}
		\nonumber & \left. C(\eta, \phi) \right|_{\textnormal{interface}}  =  \left( \sum^{'}_{\displaystyle \stackrel{\displaystyle \alpha \neq 3 \ H_{\alpha} = 0,\pm 1}{H_{3} = 0}} + \ \exp(\beta \tilde{\nu_{S}}) \sum^{'}_{\displaystyle \stackrel{\displaystyle \alpha \neq 3 \ H_{\alpha} = 0,\pm 1}{H_{3} = \pm 1}} \right) 
		\\ \nonumber & \qquad \qquad \qquad \qquad \qquad \qquad \exp\left[ i \sum_{\alpha} ( H^{2}_{\alpha}(r) {\eta}(r+e_{\alpha}) + H_{\alpha}(r) {\phi}(r+e_{\alpha}) )  \right]
		\\ \nonumber & \ \ = \ C(\eta, \phi) + \nu_{S} \sum^{'}_{\displaystyle \stackrel{\displaystyle \alpha \neq 3 \ H_{\alpha} = 0,\pm 1}{H_{3} = \pm 1}} \exp\left[ i \sum_{\alpha} ( H^{2}_{\alpha}(r) {\eta}(r+e_{\alpha}) + H_{\alpha}(r) {\phi}(r+e_{\alpha}) )  \right]
		\\ & \ \ \equiv \ C(\eta, \phi) + \nu_{S} C^{'}(\eta, \phi) \label{Caffected}
	\end{align}
where, prime over $H_{\alpha}$ sum implies constraints Eqs.(\ref{H2constraint},\ref{Hconstraint}), $C^{'}(\eta, \phi)$ corresponds to affected orientations only i.e., those with $H_{3} = \pm 1$ and, $\nu_{S} \equiv \exp(\beta \tilde{\nu_{S}}) - 1$ is a function of surface-water interaction strength. The site functional $Z_{I}$ for any interfacial site can be arranged as :
	\begin{equation}
		Z_{I}  \ = \ Z_{site} +  \nu_{S} \mu C^{'}(\eta, \phi)
		\label{interfaceaction}
	\end{equation}
By definition, $\nu_{S}$ ranges from $-1$ to $\infty$. We remark that for a perfect hydrophobic surface, $\nu_{S} = 0$.

The modified site functional at interfacial sites can be recast in the expression for full partition function, such that the following decomposition is deduced:
	\begin{align}
		\nonumber Z_{\left| \right|}  & \ = \   \int [{\cal D} \eta] [{\cal D}\phi] \prod_{r} Z_{site} \prod_{r_1 \in I_1} \left( 1 + \Gamma(r_1) \right) \prod_{r_2 \in I_2} \left( 1 + \Gamma(r_2) ) \right)
		\\   & \ = \ Z \left\langle  \exp\left( \sum_{r_1 \in I_1} \ln( 1 + \Gamma(r_1) ) + \sum_{r_2 \in I_2} \ln( 1 + \Gamma(r_2) ) \right) \right\rangle 
		\label{Zinterfaces}
	\end{align}
where, $Z_{\left| \right|}$ is partition function for the system with surfaces; $Z$ is for corresponding unperturbed case ($\nu_{S} = 0$) with $\eta = \phi = 0$ on surfaces and $\Gamma(r)$ is defined only on interfacial sites. It is relative orientational weight of affected orientations with respect to $Z_{site}$, i.e.,
	\begin{equation}
		\Gamma(r) \ = \  \frac{ \nu_{S} \mu  C^{'}(\eta, \phi) }{ Z_{site}(r) } 
		\label{affectedorientations}
	\end{equation}

The partition function for unperturbed case $Z$ can be evaluated using MMF technique. The leading mean field energy is obtained from the maximum of $Z_{site}$ at each site and fluctuations in $\eta$, $\phi$ fields are analyzed subject to vanishing boundary conditions on the surfaces. The interfaces-dependent part in $Z_{\left| \right|}$ is evaluated using cluster technique and the corresponding free energy is obtained. The resulting form of total free energy $G_{tot}$ per unit lattice area is organized to be:
	\begin{equation}
		G_{tot} = G_{o} + G_{C} + \gamma_{S_1} + \gamma_{S_2} + G_{\Gamma}
		\label{Gfparts}
	\end{equation}
where, $G_{o} + G_{C}$ is the free energy obtained from evaluation of $Z$, analogous to Eq.(\ref{oneloop}). $G_{o}$ includes leading terms proportional to $L$ and constants obtained in large $L$ limit. They contribute only to bulk pressure of the system. $G_{C}$ is the remaining $L$-dependent part. $\gamma_{S_1}$, $\gamma_{S_2}$ are free energy contributions due to surface-water interaction and evaluated only on respective interface sites $I_1$ and $I_2$ respectively. $G_{\Gamma}$ constitutes terms which involve sites of both interfaces. Expression for each of the terms is deduced in the remaining section and their relevance to hydrophobic interaction is elucidated.

We first evaluate $Z$ using the MMF technique described in the previous section. We identify the maximum of the functional to be at $\eta = \phi = 0$. It yields mean field free energy, which to the leading order is given by $L G_{m}$ (Eq.\ref{equationofstate}). The dual fields are then expanded upto quadratic order about their maximum and the resulting Gaussian functional is integrated over all field configurations to give one-loop contribution to free energy. In the process, the following Fourier transform is employed which satisfies the boundary conditions :
	\begin{equation}
		\eta(\vec{r}) \ = \ \frac{2}{L} \ \sum_{n = 1}^{L-1} \ \int\limits_{-\pi}^{\pi} \frac{(dk_1) (dk_2) }{(2 \pi)^{2}} \ \tilde{\eta}(\vec{k}) \exp(i k_1 x + i k_2 y) \sin\left( {n \pi z}/{L} \right)
		\label{FTcbc}
	\end{equation}
where $\vec{r} = (x,y,z)$ is position vector and $\vec{k} = (k_1,k_2,k_3 = \frac{n \pi}{L})$ denote modes in momentum space. Similarly for $\phi$ field.

The entropy contribution to free energy for unperturbed system is discrete analog of that of bulk water (Eq.\ref{oneloop}), in that the integral over wavevector in $z$-direction is replaced by a summation over a restricted number of wavevectors i.e., $k_3 = \frac{\pi}{L}, \frac{2 \pi}{L}, \ldots, \frac{\pi(L-1)}{L}$. To analyze $L$-dependence, we define entropy contribution per unit area in each mode in $z$-direction as :
	\begin{equation}
		S(k_{3}) \ = \ \frac{1}{2} \int\limits_{-\pi}^{\pi} \frac{ (dk_1) (dk_2) }{(2 \pi)^{2}} \ln\left(P_{\eta \eta}(\vec{k}) P_{\phi \phi}(\vec{k}) \right)
	 \label{entropy_per_mode}
	\end{equation}
where, the propagators $P_{\eta \eta}$, $P_{\phi \phi}$ are the same as in the case of bulk water. Total entropy contribution to free energy of confined water is $S(k_{3})$ summed over allowed values of $k_{3}$. Its large-$L$ behavior can be enumerated using Euler-Maclaurin series expansion \citep{Boas1971} :
	\begin{equation}
		 \sum_{k_{3} = \frac{\pi}{L}}^{\frac{\pi}{L}(L-1)} S(k_{3}) \ = \ L \int\limits_{0}^{\pi} \frac{d{k_{3}}}{\pi} S(k_{3}) - \frac{1}{2} \left( S(0) + S(\pi) \right) + \beta G_{C}
		\label{EMseries}
	\end{equation}
On right hand side of Eq.(\ref{EMseries}), first term is total entropy contribution in the same volume of bulk water. $S(0)$ and $S(\pi)$ are free energy densities in modes $k_{3} = 0$ and $k_{3} = \pi$ respectively. They are independent of $L$. From Eq.(\ref{EMseries}) we infer $G_{C}$ to be the net difference in entropy contribution per unit area between confined water and bulk water in the same volume. $G_{C}$ can be calculated as a series expansion in $\frac{1}{L}$, the leading term being : 
	\begin{equation}
		\beta G_{C} \  \simeq \ \frac{\pi}{B_{2} L} \left[ \left. \frac{\partial}{\partial k_{3}} S(k_{3}) \right|_{k_{3} = \pi} - \left. \frac{\partial}{\partial k_{3}} S(k_{3}) \right|_{k_{3} = 0} \right] \quad \textnormal{for large L}
		\label{Gcasimir}
	\end{equation}
where, $B_{2} = 2$ is first Bernoulli constant. $G_{C}$ is analogous to the Casimir interaction energy derived for the case of conducting plates confining electromagnetic fluctuations \citep{Casimir1948}. Hence, we call $G_{C}$ the \textit{Casimir part} of free energy. It falls-off asymptotically as $\frac{1}{L}$ for large $L$.

In the expression for partition function (Eq.\ref{Zinterfaces}) average over interface terms is now pursued. At each interfacial site, $\ln(1 + \Gamma(r)) \simeq \Gamma(r)$ is the leading order term. This is justified because in Eq.(\ref{affectedorientations}) for $\Gamma(r)$, we note that $\frac{\mu C^{'}(\eta,\phi)}{Z_{site}} \simeq \frac{\rho C^{'}(\eta,\phi)}{90}$ whose maximum value is always less than $1$, since $\rho < \frac{1}{3}$ and $\left| \frac{ C^{'}(\eta, \phi) }{90} \right| < \frac{2}{3}$, $C^{'}(0,0) = 60$. From Eq.(\ref{Zinterfaces}) the leading order contribution from interface terms is given by :
	\begin{equation}
		\frac{Z_{\left| \right|}}{Z} \ = \ \left\langle \exp\left( \sum_{r_{1} \in I_{1}} \Gamma(r_{1}) + \sum_{r_{2} \in I_{2}} \Gamma(r_{2}) \right) \right\rangle
	\end{equation}
The average can be evaluated using cluster technique\footnote{If $A$ and $B$ are functions of a random variable whose probability distribution is known, the average $<\exp(A + B)>$ over the probability distribution is given by : $ \left\langle \exp(A + B) \right\rangle = \exp\left[ <A> + <B> + \frac{1}{2} (<A^{2}> - <A>^{2} + <B^{2}> - <B>^{2}) + <AB> - \right.$ $ \left. <A> <B> + \ldots \right] $ }. Terms that involve sites of same interface and those involving sites of both interfaces are segregated. $\gamma_{S}$ is defined to constitute terms corresponding to sites on same interface. Each of them is proportional to $\nu_{S}$ or its higher order. $\gamma_{S}$ is given to the leading order as :
	\begin{equation}
		{- \beta \gamma_{S} A}{} \ = \  \left[ \left\langle \sum_{r \in I}  \Gamma(r) \right\rangle + \left\langle \sum_{\stackrel{r_{1}, r_{2} \in I}{r_{1} \neq r_{2}}}  \Gamma(r_{1})  \Gamma(r_{2})  \right\rangle - \left\langle \sum_{r \in I}  \Gamma(r)  \right\rangle^{2} \right] \quad \label{interfacialtension}
	\end{equation}
where, $A$ is area of the surface. $\gamma_{S}$ arises due to surface-water interaction and consequent effect on orientational fluctuations in the interfacial region. 

Each of the averages in Eq.(\ref{interfacialtension}) can be evaluated using a functional integration relation\footnote{ If $\phi$ is a random field whose action is known and when a constant external field $J$ couples to $\phi$ such that their interaction is $i J \phi(r)$, then $< \exp(i J (\phi(r_{1}) + \phi(r_{2})) ) >$  $= \exp[- \frac{1}{2} J^{2} ( < \phi(r_{1}) \phi(r_{1}) > + <\phi(r_{2}) \phi(r_{2}) + 2 < \phi(r_{1}) \phi(r_{2}) > ) + \ldots ]$. If two-point correlation is the leading order, then the subsequent terms of higher order denoted by $(\ldots)$ can be ignored}. For an interface site with surface in $e_{3}$ direction, using Eqs.(\ref{Caffected},\ref{affectedorientations}) $\left\langle \Gamma(r) \right\rangle $ is given to the leading order as :
	\begin{align} 
		\nonumber \left\langle \Gamma(r) \right\rangle \  = \ \left( \frac{\nu_{S} \rho}{90} \right) \sum^{'}_{\displaystyle \stackrel{\displaystyle \alpha \neq 3 \ H_{\alpha} = 0,\pm 1}{H_{3} = \pm 1}}  \exp & \left[ \sum_{\alpha, \alpha^{'}} \left( H^{2}_{\alpha}(r) H^{2}_{\alpha^{'}}(r) {\cal G}_{\eta}(r+e_{\alpha}, r+e_{\alpha^{'}}) \right. \right. 
		\\ & \left. \left.  + \ H_{\alpha}(r) H_{\alpha^{'}}(r) {\cal G}_{\phi}(r+e_{\alpha}, r+e_{\alpha^{'}}) \vphantom{H^{2}_{\alpha^{'}}} \right) \vphantom{\sum_{\alpha}} \right]  \label{average_Gamma}
	\end{align}	
where, the $H_{\alpha}$ summation is over affected orientations at site $r$. The prime over summation indicates $H_{\alpha}$'s of each orientation satisfy Eqs.(\ref{H2constraint},\ref{Hconstraint}). The exponential in Eq.(\ref{average_Gamma}) corresponds to one such orientation. $H_{\alpha}$, $H_{\alpha^{'}}$ are bond arms of the same orientation; $r + e_{\alpha}$, $r + e_{\alpha^{'}}$ are the bond arm locations. The average $\left\langle \Gamma(r_{1}) \Gamma(r_{2}) \right\rangle$ is given to leading order as :
	\begin{align}
		\nonumber & \left\langle \Gamma(r_{1}) \Gamma(r_{2}) \right\rangle \ = \ \displaystyle  \left( \frac{\nu_{S}  \rho}{90} \right)^{2} 
		\sum^{'}_{\displaystyle \stackrel{\displaystyle \alpha \neq 3 \ H_{\alpha} = 0,\pm 1}{H_{3} = \pm 1}} \ \  \sum^{'}_{\displaystyle \stackrel{\displaystyle \kappa \neq 3 \ H_{\kappa} = 0,\pm 1}{H_{3} = \pm 1}}  \exp \left[ \vphantom{\sum_{\alpha^{'}}} \right.
		\\ \nonumber & \left.  \displaystyle \sum_{\alpha, \alpha^{'}} \left( H^{2}_{\alpha}(r_{1}) H^{2}_{\alpha^{'}}(r_{1}) {\cal G}_{\eta}(r_{1} + e_{\alpha}, r_{1} + e_{\alpha^{'}}) + H_{\alpha}(r_{1}) H_{\alpha^{'}}(r_{1}) {\cal G}_{\phi}(r_{1} + e_{\alpha}, r_{1} + e_{\alpha^{'}}) \right) \right.
		\\ \nonumber & \left. \displaystyle  + \sum_{\kappa, \kappa^{'}} \left( H^{2}_{\kappa}(r_{2}) H^{2}_{\kappa^{'}}(r_{2}) {\cal G}_{\eta}(r_{2} + e_{\kappa}, r_{2} + e_{\kappa^{'}}) + H_{\kappa}(r_{2}) H_{\kappa^{'}}(r_{2}) {\cal G}_{\phi}(r_{2} + e_{\kappa}, r_{2} + e_{\kappa^{'}}) \right) \right.
		\\ & \left. \displaystyle  + \sum_{\alpha, \kappa} \left( H^{2}_{\alpha}(r_{1}) H^{2}_{\kappa}(r_{2}) {\cal G}_{\eta}(r_{1} + e_{\alpha}, r_{2} + e_{\kappa}) + H_{\alpha}(r_{1}) H_{\kappa}(r_{2}) {\cal G}_{\phi}(r_{1} + e_{\alpha}, r_{2} + e_{\kappa}) \right)  \vphantom{\sum_{\alpha^{'}}}  \right] \label{average_GammaGamma} 
	\end{align}
where, $H_{\alpha}$, $H_{\alpha^{'}}$ are bond arms of an affected orientation at site $r_{1}$ and $H_{\kappa}$, $H_{\kappa^{'}}$ are those of an orientation at site $r_{2}$. The exponential corresponds to product of the two orientations and the summation is over all possible products. The two-point Green's function ${\cal G}_{\eta}(r_{1},r_{2})$ for $\eta$-field fluctuations between any two arbitrary sites $r_{1} = (x_{1},y_{1},z_{1})$ and $r_{2} = (x_{2},y_{2},z_{2})$ is given by, 
	\begin{align}
		{\cal G}_{\eta}(r_{1}, r_{2}) \  = \ \frac{2}{L} \ \sum_{n = 1}^{L-1} \ \int\limits_{-\pi}^{\pi} \frac{ (dk_{1})(dk_{2}) }{(2 \pi)^{2}} \exp\left(i k_{1} (x_{1} -  x_{2}) + i k_{2} (y_{1} - y_{2}) \right)   \frac{\displaystyle  \sin\left( {n \pi z_{1}}/{L}\right) \sin\left( {n \pi z_{2}}/{L}\right) }{ P_{\eta \eta}(\vec{k}) }  \label{Greensfunction}
	\end{align} 
Similarly, ${\cal G}_{\phi}(r_{1}, r_{2})$ for $\phi$ field can be defined using the propagator $P_{\phi \phi}(\vec{k})$.

The expression for $\gamma_{S}$ indicates that it varies with separation distance, owing to the $L$-dependent Green's functions. The asymptotic value of $\gamma_{S}$ is the interfacial tension for hydrophobic surface in contact with water. The leading correction term is proportional to $\frac{1}{L}$ for large-$L$ and contributes to force between the surfaces.

From the cluster expansion of partition function, terms that involve sites of both interfaces are grouped as $G_{\Gamma}$. It is given to the leading order as :
	\begin{equation}
		{- \beta G_{\Gamma} A } \ = \ \left[ \left\langle \sum_{r_{1} \in I_{1}}  \Gamma(r_{1}) \sum_{r_{2} \in I_{2}}  \Gamma(r_{2})  \right\rangle - \left\langle \sum_{r_{1} \in I_{1}}   \Gamma(r_{1})  \right\rangle  \left\langle \sum_{r_{2} \in I_{2}}  \Gamma(r_{2})  \right\rangle  \right] \label{IIFE}
	\end{equation}
Effectively, $G_{\Gamma}$ is connected correlation between orientational fluctuations of both interfaces. Hence, we call this contribution \textit{interfacial fluctuations-induced part} of free energy. The averages in Eq.(\ref{IIFE}) can be evaluated using Eq.(\ref{average_Gamma}) with $\nu_{S}$ corresponding to each interface and using Eq.(\ref{average_GammaGamma}) with proportionality factor $(\nu_{S_{1}} \nu_{S_{2}})$ instead of $(\nu_{S})^{2}$. The identity of sites is as per given in the expression for $G_{\Gamma}$ (Eq.\ref{IIFE}).

The long distance behavior of $G_{\Gamma}$ is dominated by $\phi(r)$ correlations, $\eta(r)$ being short ranged. Between two hydrophobic surfaces, to the leading order $G_{\Gamma}$ is proportional to square of orientational correlations i.e., $({\cal G}_{\phi}(r))^{2}$, where ${\cal G}_{\phi}(r)$ is an exponentially falling-off function for large $r$ (Appendix \ref{app_correlationfunctions}). 

For the case of mesoscopic surfaces hydrophobic force is suggested to arise from orientational correlations between water molecules at both interfaces \citep{KanthPRE2010}. The force is seen to decay exponentially with separation distance, asymptotically. $G_{\Gamma}$ is thus analogous to hydrophobic interaction free energy of mesoscopic surfaces. However for macroscopic surfaces, in addition to $G_{\Gamma}$, hydrophobic force obtains contributions from Casimir part and interfacial tension. This distinguishes hydrophobic interaction between large surfaces from that of between small surfaces both qualitatively and quantitatively. The non-additive nature of hydrophobic interaction with increasing size of surfaces has attracted considerable attention \citep{ChandlerNature2005,*AshbaughRMP2006} and our work provides a direction to elucidate the size dependence in terms of hydrogen bond fluctuations in water.

\subsection{Hydrophilic surfaces} {\label{section_hydrophilicsurfaces}}

We can envisage surfaces of generic heterogeniety in our calculation. The heterogeniety could be in terms of space-dependent $\nu_{S}$ and/or charge on surface. One of the simplest cases is a homogeneous hydrophilic surface with a fixed charge on each site. We first consider the case of a positively charged hydrophilic surface. On its interface, the site functional comprises weights corresponding to all states. When a water molecule is present on interface, its hydrogen arm is restricted from pointing in surface direction. We assign an energetic penalty to such orientations and the site functional can be arranged, analogous to the case of a hydrophobic surface, as :
 	\begin{equation*}
		Z_{I} \ = \ Z_{site} +  \nu_{S} \mu C^{'}(\eta, \phi)
	\end{equation*}
Here, $\nu_{S} \in (-1,0)$ (ideally, $\nu_{S} = -1$) and the orientational weights corresponding to affected orientations $C^{'}(\eta, \phi)$ are given by :
	\begin{equation}
		C^{'}_{+}(\eta, \phi)  =  \sum^{'}_{\displaystyle \stackrel{\displaystyle \alpha \neq 3 \ H_{\alpha} = 0,\pm 1}{H_{3} = 1}} \exp\left[ i \sum_{\alpha} ( H^{2}_{\alpha}(r) {\eta}(r+e_{\alpha}) + H_{\alpha}(r) {\phi}(r+e_{\alpha}) )  \right]
		\label{Caffected_pos_hydrophilic}
	\end{equation}
The above expression is for an interface site with surface in $e_{3}$ direction. A negatively charged hydrophilic surface can also be envisaged such that for interface water orientations with lone-pair arm in surface direction are energetically penalized. Here, the weights for affected orientations are :
	\begin{equation}
		C^{'}_{-}(\eta, \phi)  =  \sum^{'}_{\displaystyle \stackrel{\displaystyle \alpha \neq 3 \ H_{\alpha} = 0,\pm 1}{H_{3} = -1}} \exp\left[ i \sum_{\alpha} ( H^{2}_{\alpha}(r) {\eta}(r+e_{\alpha}) + H_{\alpha}(r) {\phi}(r+e_{\alpha}) )  \right]
		\label{Caffected_neg_hydrophilic}
	\end{equation}

We now compute the free energy components $G_{C}$, $\gamma_{S_{1}}$, $\gamma_{S_{2}}$, $G_{\Gamma}$ using their respective expressions for different types of surfaces. $\nu_{S}$ is an arbitrary parameter in the calculation. It is chosen close to its ideal value for each surface type. The properties of water enter the computation via Green's functions ${\cal G}_{\eta}$, ${\cal G}_{\phi}$. These are computed within the model using Eq.(\ref{Greensfunction}). Due to $L$-dependent modes in the confined direction, all the free energy components that depend on fluctuations are expected to vary with separation distance $L$.

\section{Results : Hydrophobic force, interfacial tension} {\label{section_hydrophobicforce}}

\begin{figure*}
		\epsfig{file=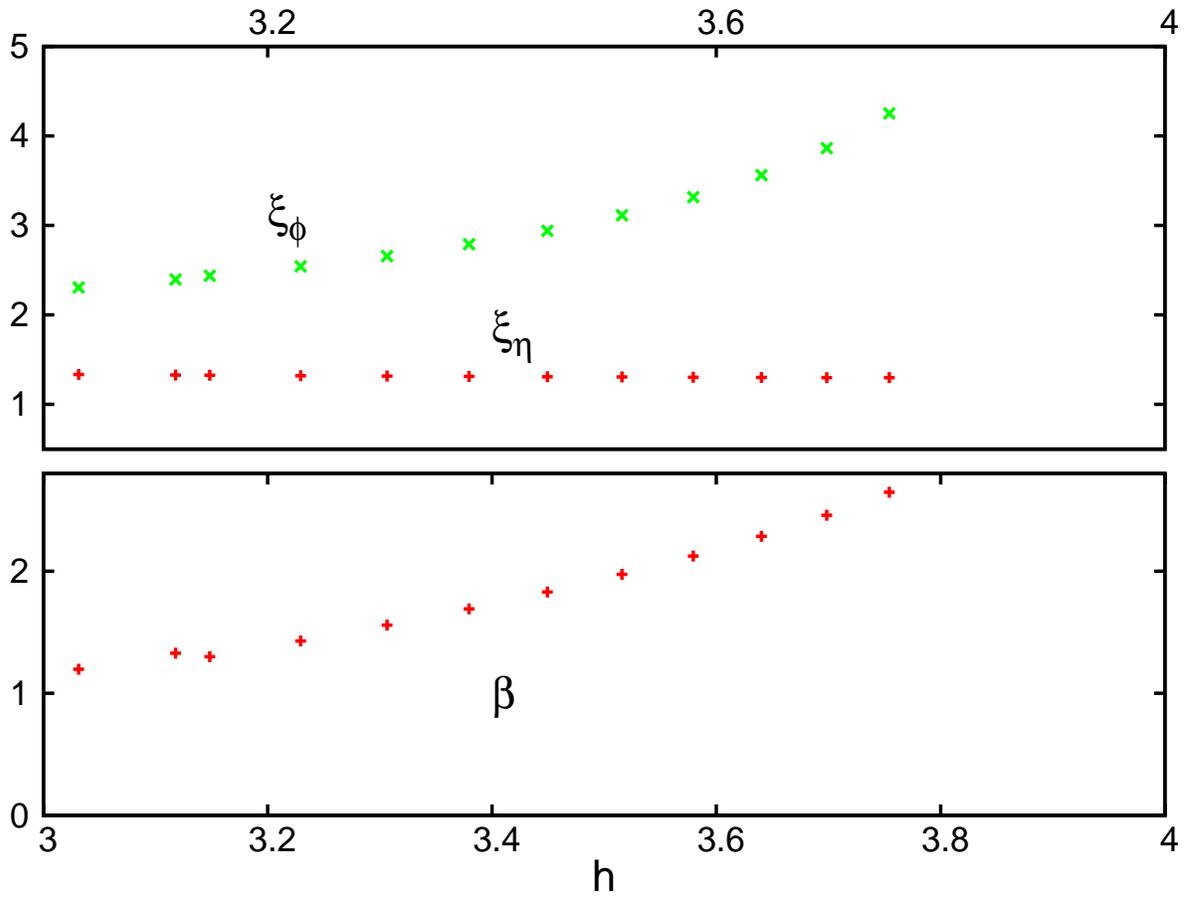,width=1.0\linewidth,height=0.5\textheight}
		\caption{\label{fig_h_beta} (Color online) Inverse temperature $\beta$ and correlation lengths $\xi_{\eta}$, $\xi_{\phi}$ as a function of $h$. $\beta$ is measured in units of hydrogen bond strength $\tilde{\lambda}$ in the model. Correlation lengths are expressed in lattice units.}
	\end{figure*}

We first mention that this computation is totally parameter-free on lattice. Hence, the best way to interpret results is in terms of physically observable quantities such as $\rho$ and hydrogen bond density. Indeed because of equation of network (Eq.\ref{equationofnetwork}) only one of them is independent. We find that it is best to describe in terms of $h = \frac{2 \textnormal{HB}}{\rho}$, the average number of hydrogen bonds per molecule. Temperature is conjugate to HB (total hydrogen bond density) and hence it is also implicitly fixed self-consistently due to equation of network, as shown in Fig.(\ref{fig_h_beta}). The relation between $\rho$ and $h$ is simple at zeroth order in MMF theory (Eq.\ref{equationofnetwork_h_rho}), but it becomes non-linear at one-loop level. (Zeroth order is still a reasonable approximation \citep{KanthPhysica2011}.) Hence, all densities DB, HB, $\rho$ and the free energy components given by Eqs.(\ref{EMseries},\ref{interfacialtension},\ref{IIFE}) are evaluated from partition function upto one-loop order using the corresponding expressions for propagators (Appendix \ref{app_correlationfunctions}).

In our model MMF theory describes liquid for $h > 3$ reasonably consistently. For $2 < h < 3$ MMF approximation is not seen to be good, namely, one-loop order terms are either comparable or exceed zero-loop term. So, we choose to present our results for $h > 3$. All the potentials and energies are computed in the units of hydrogen bond strength $\tilde{\lambda}$ taken to be unity. The lattice constant in the model is arbitrary. By computing physical lengthscales such as correlation length it can be fixed. Correlation lengths for density ($\xi_{\eta}$) and orientational fluctuations ($\xi_{\phi}$) to the leading order are simple expressions given in Appendix (\ref{app_correlationfunctions}), but a precise expression to one-loop order is implicitly given. In Fig.(\ref{fig_h_beta}) we plot correlation lengths as a function of $h$. $\xi_{\eta}$ is only about one lattice unit in liquid phase and does not vary considerably with $h$, while $\xi_{\phi}$ increases with $h$. In MD simulation density correlation length is not seen; this is consistent with MMF result since $\xi_{\eta}$ is equal to the minimum length possible in the model and also independent of $h$. Orientational correlation lengths inferred from MD simulation are $5.2$ \AA{} and $24$ \AA{}, of which the latter is weaker in strength (one-tenth) relative to the shorter one \citep{KanthPRE2010}. In our water model we have only one orientational correlation length $\xi_{\phi}$ which we relate to $5.2$ \AA{}. For liquid water $h$ value is suggested to be about $3.6$ \citep{MahoneyJCP2000Vol112}. From Fig.(\ref{fig_h_beta}) $h = 3.58$ corresponds to $\xi_{\phi} \simeq 3.3$ lattice units. Consequently, we infer that $1$ lattice unit $\simeq \frac{5.2}{3.3} = 1.57$ \AA{}.

\begin{figure*}
		\epsfig{file=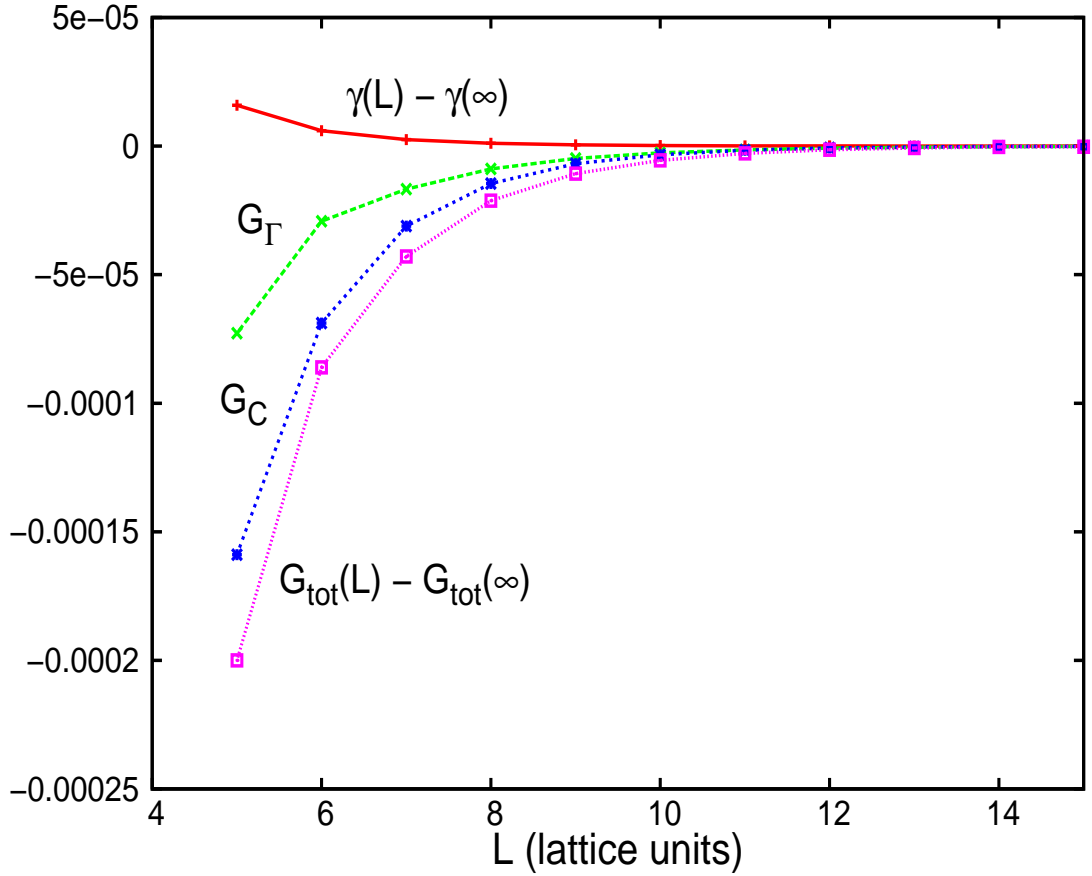,width=0.9\linewidth,height=0.5\textheight}
		\caption{\label{fig_Gall_zz} (Color online) Different contributions to $G_{tot}$ for two hydrophobic surfaces ($\nu_{S_{1}} = \nu_{S_{2}} = -0.5$). The curves are plotted for $h = 3.58$. In the order from top to bottom the curves correspond to $\gamma_{S}(L) - \gamma_{S}(\infty)$, $G_{\Gamma}$, $G_{C}$ and $G_{tot}(L) - G_{tot}(\infty)$ respectively. The free energy densities are measured per unit hydrogen bond strength.}
	\end{figure*}

In Fig.(\ref{fig_Gall_zz}) various contributions to interaction free energy and their relative magnitudes are plotted as a function of separation distance $L$ between surfaces. The plot is presented for $h = 3.58$. Casimir part $G_{C}$  gives the most attractive force, followed by $G_{\Gamma}$, while the interfacial term $\gamma_{S}$ is repulsive, albeit very small. $G_{C}$, $\gamma_{S}$ fall-off as $\frac{1}{L}$ for large $L$ from our analytic calculations. Numerically, beyond $15$ lattice units they are insignificant. All the plots are presented for lattice distance $L \geq 5$. For smaller $L$ the results are predominantly influenced by surface effects. In the model, for $L =  4$ there is only one layer which can have free orientations (besides two interface layers), while for $L \geq 5$ there are two or more such layers.

\begin{figure*}
		\epsfig{file=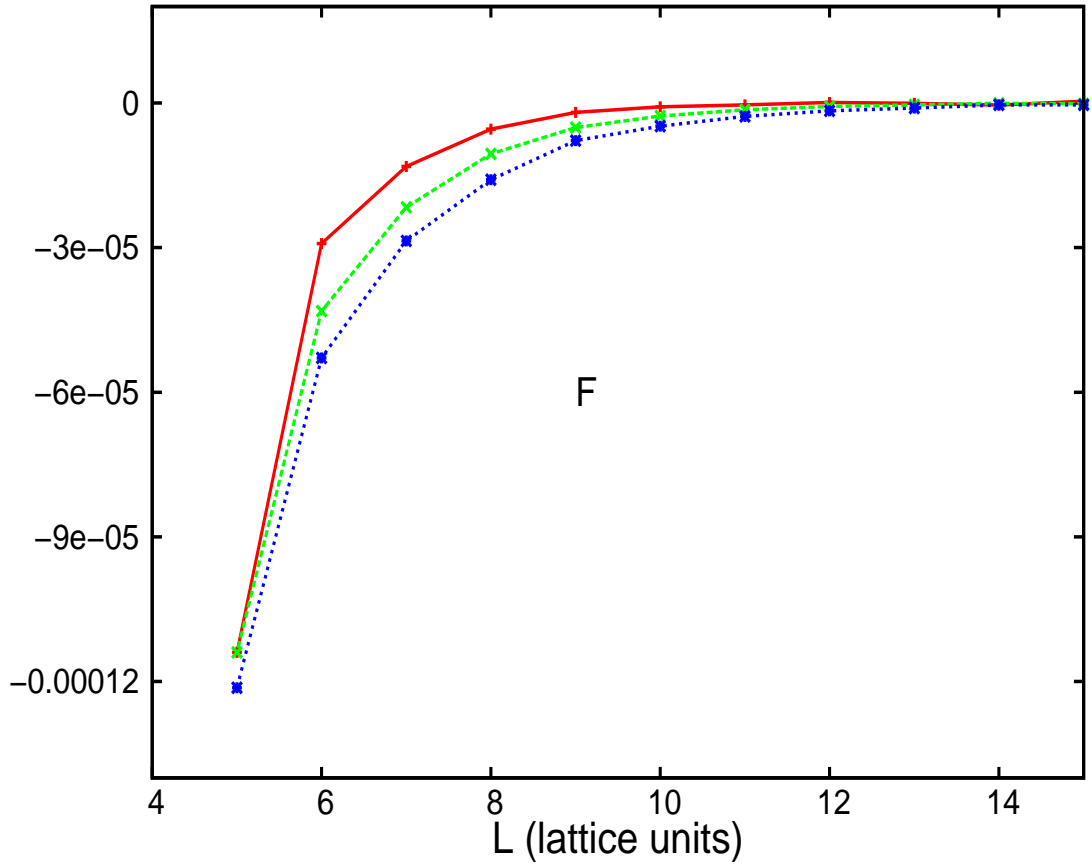,width=0.9\linewidth,height=0.5\textheight}
		\caption{\label{fig_Force_zz} (Color online) Force between two hydrophobic surfaces ($\nu_{S_{1}} = \nu_{S_{2}} = -0.5$). Top (red) curve corresponds to $h = 3.03$, middle (green) curve : $h = 3.58$, bottom (blue) : $h = 3.75$. Force is measured per unit hydrogen bond strength per unit lattice distance.}
	\end{figure*}

Force is computed as discrete derivative of total free energy with respect to $L$ and plotted in Fig.(\ref{fig_Force_zz}) for various $h$. The curves effectively show that the force can manifest upto a length of about $15$ lattice units which translates to about four times the orientational correlation length in the model. All the free energy components and force obtain major contributions from orientational fluctuations.

\begin{figure*}[h]
		\epsfig{file=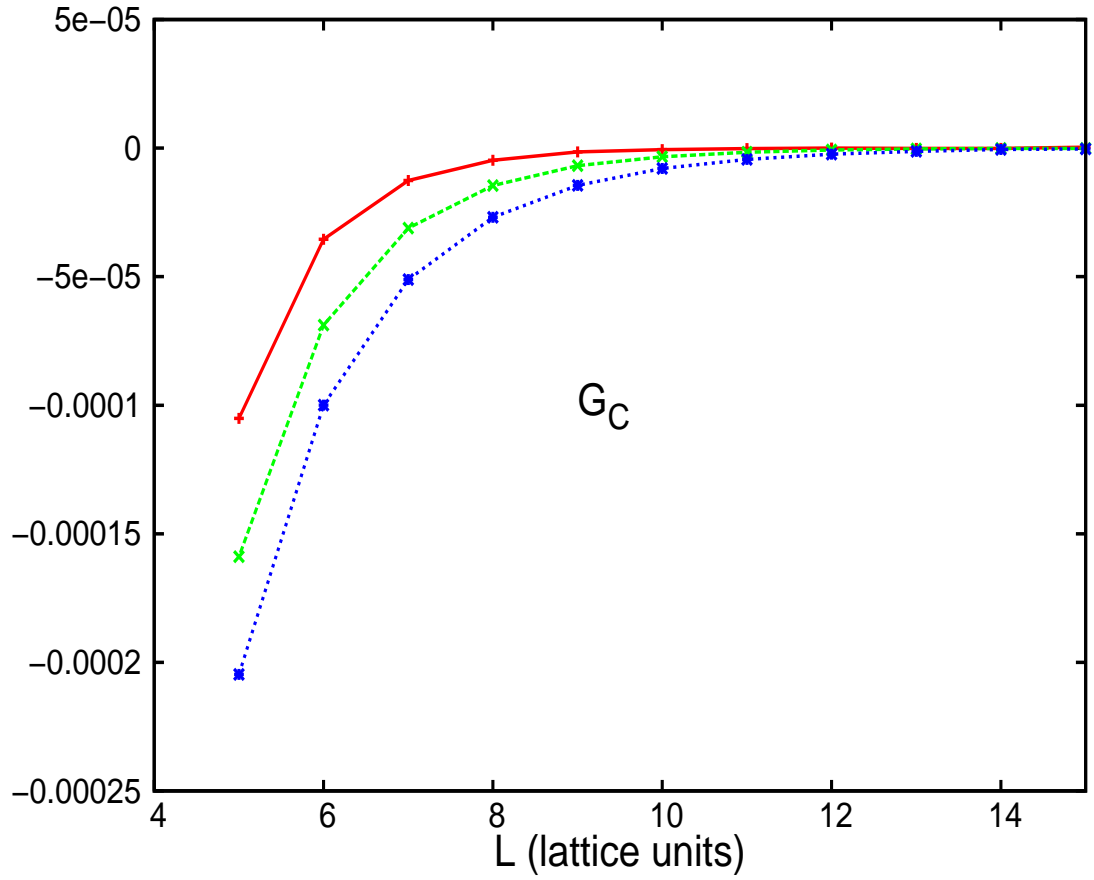,width=0.9\linewidth,height=0.5\textheight}
		\caption{\label{fig_Gcasimir} (Color online) $G_{C}$ as a function of $L$. Top (red) curve corresponds to $h = 3.03$, middle (green) curve : $h = 3.58$, bottom (blue) : $h = 3.75$. }
	\end{figure*}

\begin{figure*}
		\epsfig{file=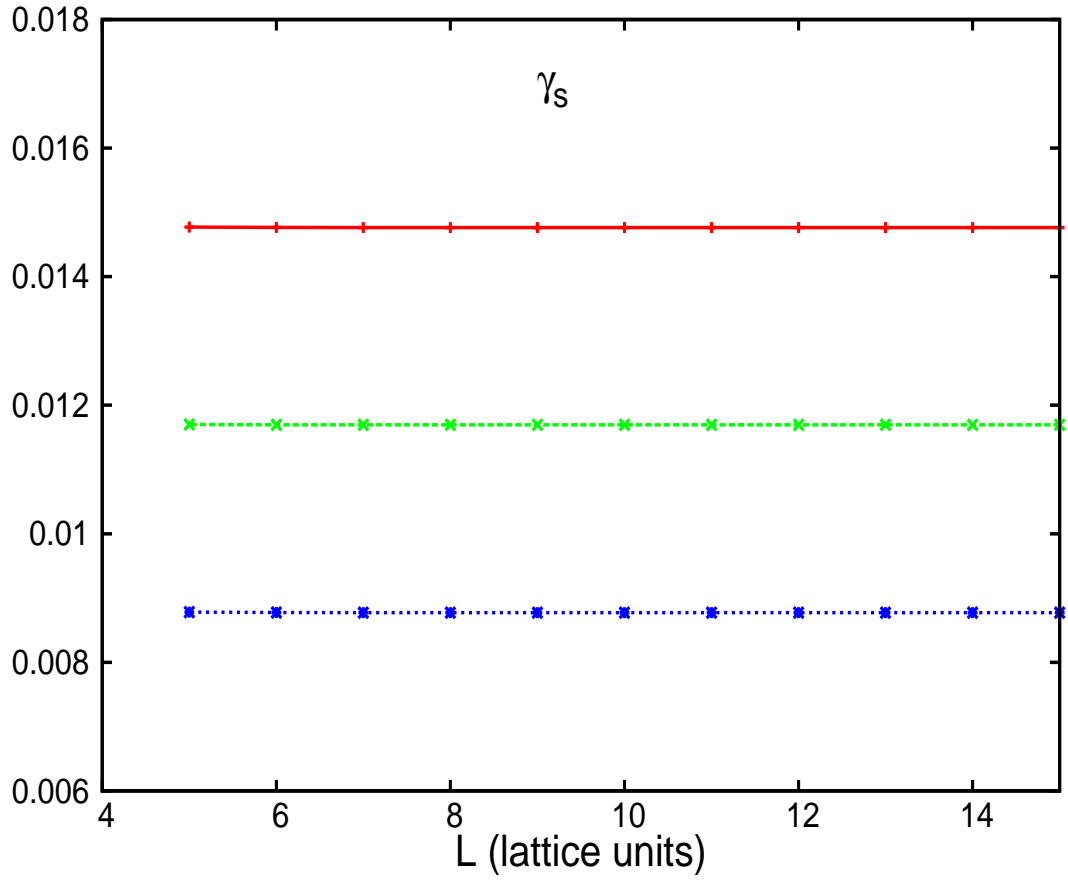,width=0.9\linewidth,height=0.5\textheight}
		\caption{\label{fig_Git_z} (Color online) $\gamma_{S}$ for a hydrophobic surface ($\nu_{S} = -0.5$). Top (red) curve corresponds to $h = 3.03$, middle (green) curve : $h = 3.30$, bottom (blue) : $h = 3.58$.}
	\end{figure*}

\begin{figure*}
		\epsfig{file=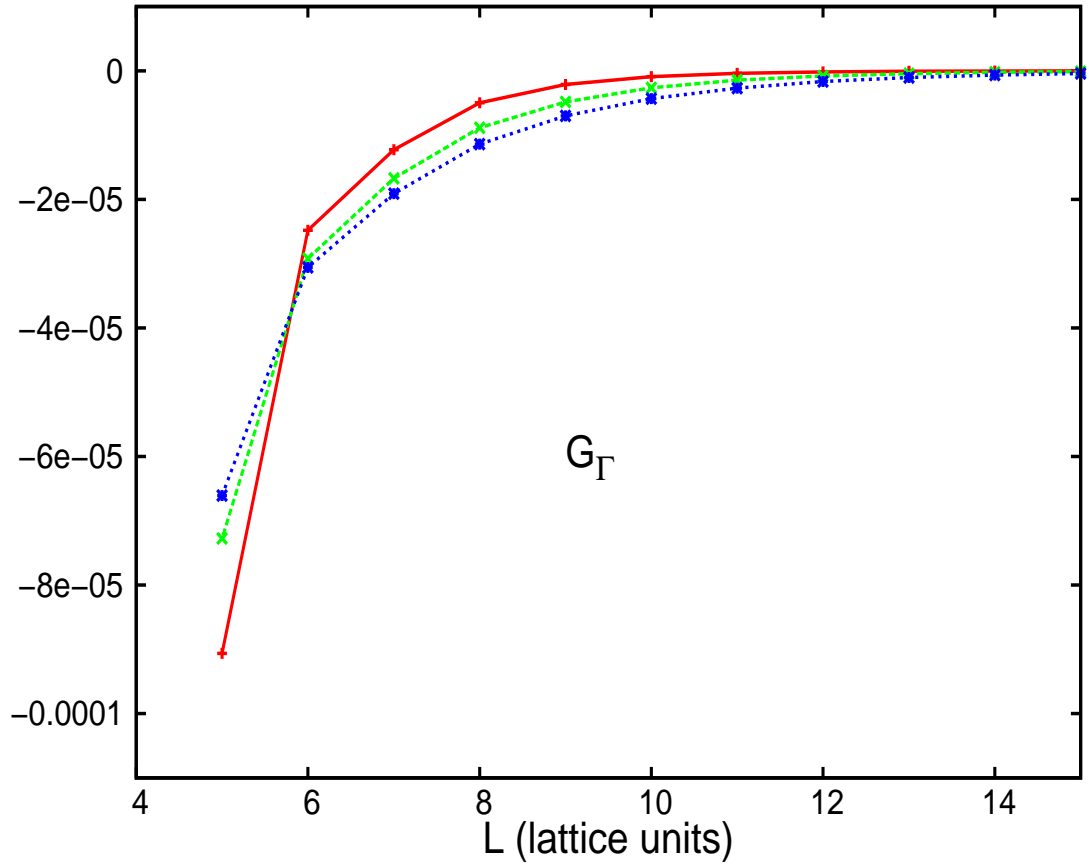,width=0.9\linewidth,height=0.5\textheight}
		\caption{\label{fig_Gf_zz} (Color online) $G_{\Gamma}$ for two hydrophobic surfaces ($\nu_{S_{1}} = \nu_{S_{2}} = -0.5$). For $L \geq 6$, top (red) curve corresponds to $h = 3.03$, middle (green) curve : $h = 3.58$, bottom (blue) : $h = 3.75$.}
	\end{figure*}

Figs.(\ref{fig_Gcasimir},\ref{fig_Git_z},\ref{fig_Gf_zz}) display the $h$-dependence of $G_{C}$, $\gamma_{S}$ and $G_{\Gamma}$ functions. The Casimir part $G_{C}$ monotonically increases in magnitude with $h$. The interfacial tension contribution $\gamma_{S}$ decreases and is always slightly repulsive. Interfacial fluctuations-induced part $G_{\Gamma}$ increases with $h$ for $L \geq 6$. At shorter distances it decreases with increasing $h$. This indicates that the adhesion strength of $G_{\Gamma}$ component is higher for higher temperatures. This behavior is qualitatively similar to the temperature dependence of interaction free energy for mesoscopic hydrophobic surfaces \citep{Frank1945,WidomPCCP2003}. This reaffirms our interpretation that $G_{\Gamma}$ component is analogous to hydrophobic interaction free energy for mesoscopic surfaces.

\begin{figure*}
		\epsfig{file=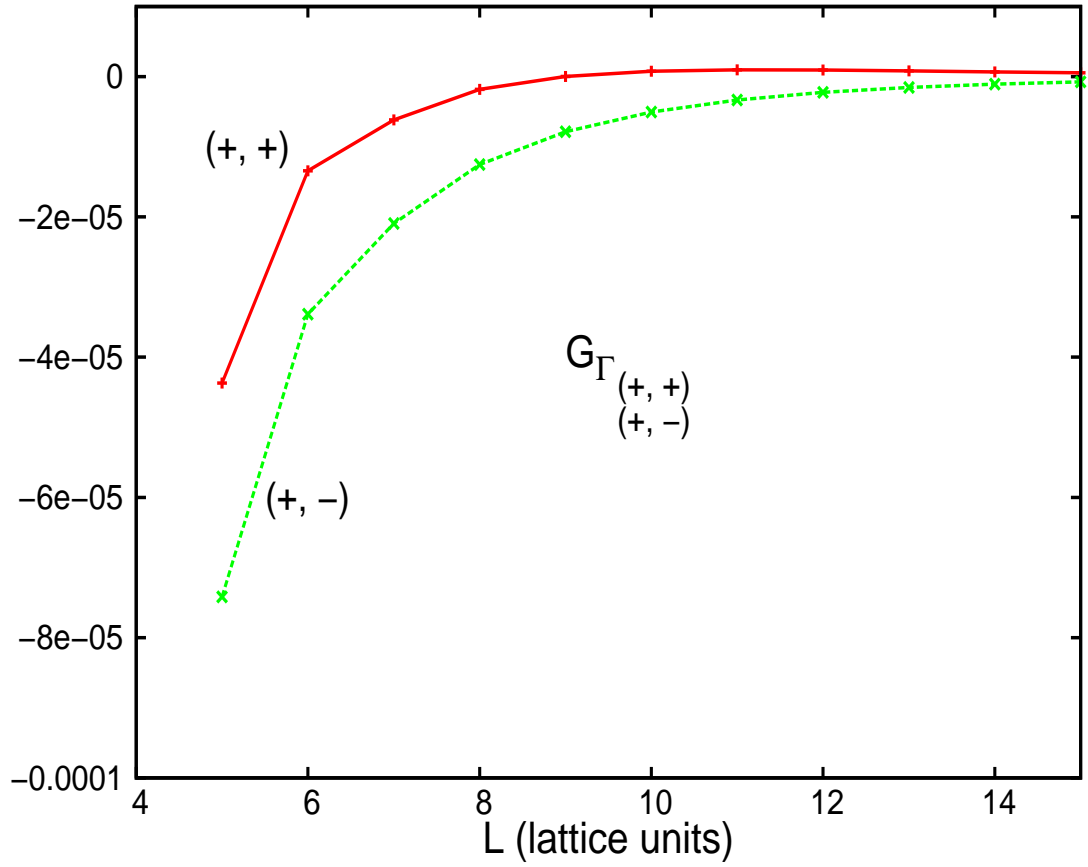,width=0.9\linewidth,height=0.5\textheight}
		\caption{\label{fig_Gf_pp_pn} (Color online) $G_{\Gamma}$ for hydrophilic surfaces ($\nu_{S_{1}} = \nu_{S_{2}} = -0.9$). $(+,+)$ curve corresponds to similar type of hydrophilic surfaces and $(+,-)$, to dissimilar type. Both curves are plotted for $h = 3.58$. }
	\end{figure*}

\begin{figure*}
	\epsfig{file=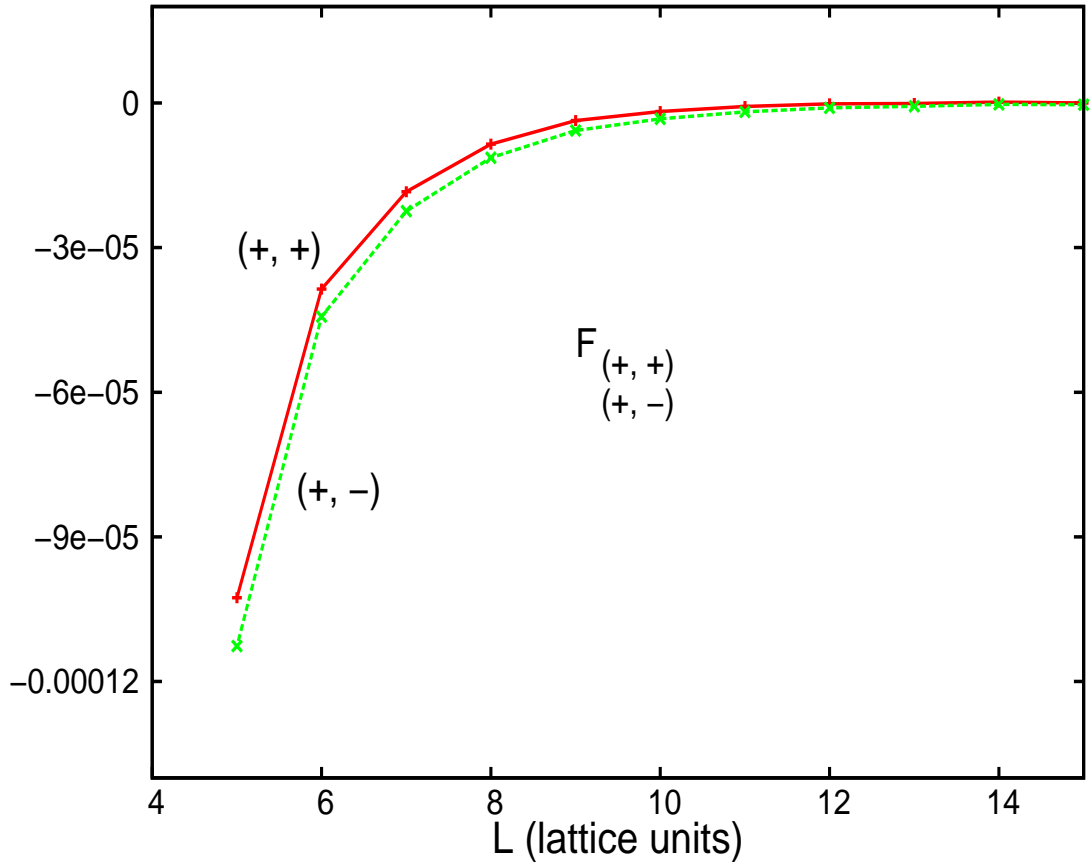,width=0.9\linewidth,height=0.5\textheight}
		\caption{\label{fig_Force_pp_pn} (Color online) Force between two hydrophilic surfaces ($ \nu_{S_{1}} = \nu_{S_{2}} = -0.9$). $(+,+)$ indicates similar type of hydrophilic surfaces and $(+,-)$ indicates dissimilar type. Both curves correspond to $h = 3.58$. }
	\end{figure*}

Fig.(\ref{fig_Gf_pp_pn}) is the plot for $G_{\Gamma}$ contribution between two hydrophilic surfaces, both of same type (hydrogen donor/acceptor) and of dissimilar type. $G_{\Gamma}$ in this case is proportional to ${\cal G}_{\phi}$ and hence, correlation length is twice as longer in range than in the case of hydrophobic surfaces (where $G_{\Gamma}$ is proportional to $({\cal G}_{\phi})^{2}$). At short distances it is seen to be attractive for both combinations. However, for large distances it is weakly repulsive between like-charged surfaces, in contrast to attraction between oppositely charged surfaces. Fig.(\ref{fig_Force_pp_pn}) depicts the force between hydrophilic surfaces for both similar and dissimilar combinations. As expected, the dissimilar pair of surfaces have marginally larger attraction than that of similar surfaces. It is interesting to note that like-charged hydrophilic surfaces also have a net attraction. This is due to dominance of Casimir part $G_{C}$ which is indifferent to surface charge.

\begin{figure*}
	\epsfig{file=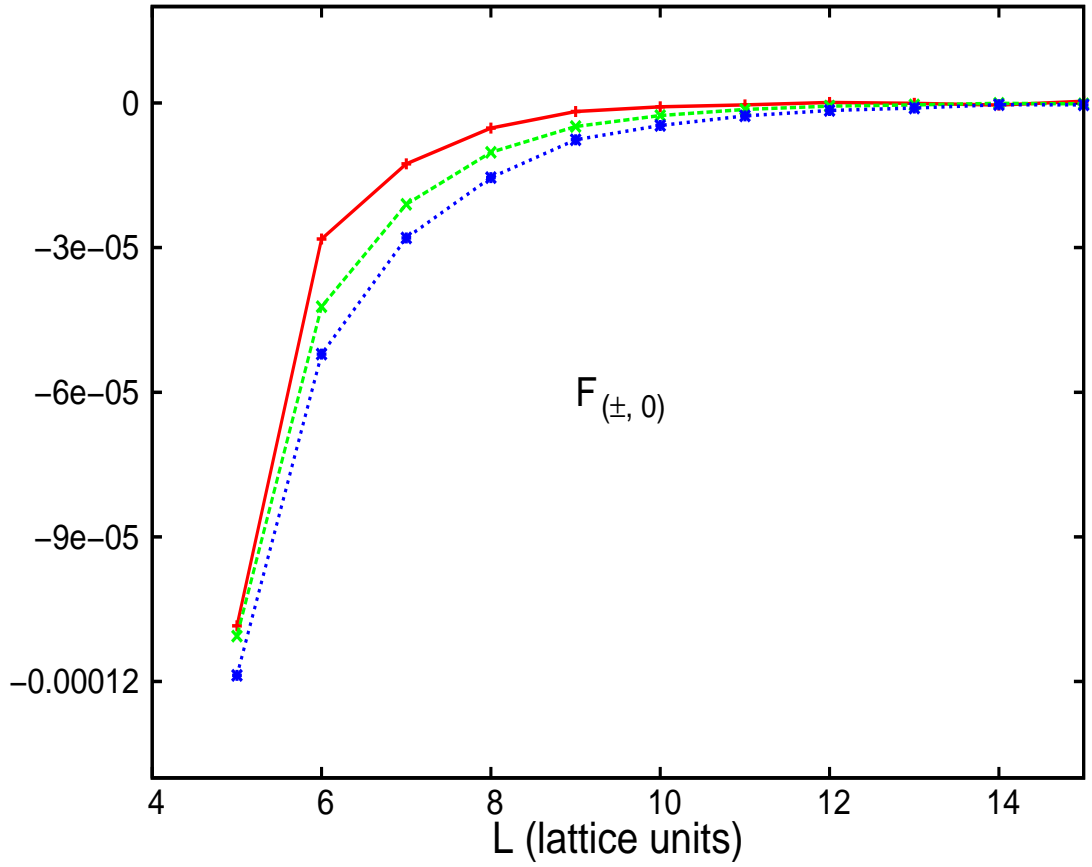,width=0.9\linewidth,height=0.5\textheight}
		\caption{\label{fig_Force_pz} (Color online) Force between a hydrophilic ($\nu_{S_{1}} = -0.9$) and a hydrophobic surface ($\nu_{S_{2}} = -0.5$). Top (red) curve corresponds to $h = 3.03$, middle (green) curve : $h = 3.58$, bottom (blue) : $h = 3.75$. $(\pm, 0)$ indicates that force is between a positively (negatively) charged hydrophilic surface and a hydrophobic surface. }
	\end{figure*}

Fig.(\ref{fig_Force_pz}) displays force between a hydrophobic and hydrophilic surface. It bears similar profile as in the case of two hydrophobic surfaces. This is expected because essentially $G_{\Gamma}$ is qualitatively same for both cases i.e., proportional to $({\cal G}_{\phi})^{2}$. For all surface combinations the force is seen to increase in magnitude with $h$, dominantly due to indifference of Casimir part to surface types. This is a consequence of the fact that the entropy induced forces are largely charge neutral.

Next, we make an attempt to relate our computational results to those of experiments. The free energy values presented in the graphs are in the units where hydrogen bond strength is unity. Generally, dimensionful quantities in lattice models and those in corresponding continuum models are not the same. So it is best to compare dimensionless quantities. In our instance, for $h = 3.58$ and $L = 6$ lattice units which translates to $6 \times 1.57 \textnormal{\AA{}} \simeq 9.5 \textnormal{\AA{}}$, $\displaystyle \frac{ |G_{tot}(6) - G_{tot}(\infty) | }{ \gamma_{S}(\infty) } \simeq \frac{9 \times 10^{-5}}{8.5 \times 10^{-3}} \simeq 10^{-2}$. From experiments interaction free energy estimate when two hydrophobic plates are about $10$ \AA{} apart is about $1 \ \textnormal{mJ} \ \textnormal{m}^{-2}$ \citep{Hammer2010}, while interfacial tension is in the range $50 - 100 \ \textnormal{mJ} \ \textnormal{m}^{-2}$ \citep{Goebel1997}; their ratio agrees with our computation. In experiments the free energy values are also measured for larger distances all the way up to $100$ \AA{}. Unfortunately our model is not good for these distances. This discrepancy was already noticed when our results were compared with MD simulation. The simple water model has only one orientational correlation length, while there are more than one in both MD simulations \citep{KanthPhysica2011} and surface force apparatus experiments \citep{Claesson2001}. We conclude that while order of magnitude estimate of the strength of hydrophobic force is in agreement with Casimir-like energies envisaged here, a few more important details are perhaps missing in our simple model of water.

\section{Transverse density profile} {\label{section_densityprofile}}

We also deduce expression for water density profile along the confinement direction $z$. $\rho(z)$ is obtained by assuming chemical potential of water $\tilde{\mu}$ to be $z$-dependent and then, a partial derivative of $\ln(Z_{||})$ is taken with respect to $\beta \tilde{\mu}(z)$. At both interfaces i.e., $z = 1$ and $z = L-1$, the modified fugacity provides additional correction to average density. The expression for density profile is given by : 
	\begin{equation}
		\rho(z)  \ \equiv \  \frac{ \partial (\ln (Z_{||})) }{ \partial (\beta \tilde{\mu}(z)) }
		\ = \ \rho_{C}(z) +  \frac{1}{A} \left\langle \frac{ \partial }{ \partial (\beta \tilde{\mu}(z)) } \left( \sum_{r_{1} \in I_{1}}  \Gamma(r_{1}) + \sum_{r_{2} \in I_{2}}  \Gamma(r_{2}) \right) \right\rangle + \ldots 
		\label{rhoz}
	\end{equation}
$\rho_{C}$ is obtained from differentiating $Z$ in Eq.(\ref{Zinterfaces}). It is the density profile between ideal hydrophobic surfaces ($\nu_{S} = 0$) and is the dominant contribution at all positions. The explicit expression for $\rho_{C}(z)$ upto one-loop order is given in the Appendix (\ref{app_densityprofile}). The interfaces-dependent term in Eq.(\ref{rhoz}) can be analyzed using Eq.(\ref{average_Gamma}). This contribution is only at $z =1$ and $z = L-1$. 

\begin{figure*}
		\epsfig{file=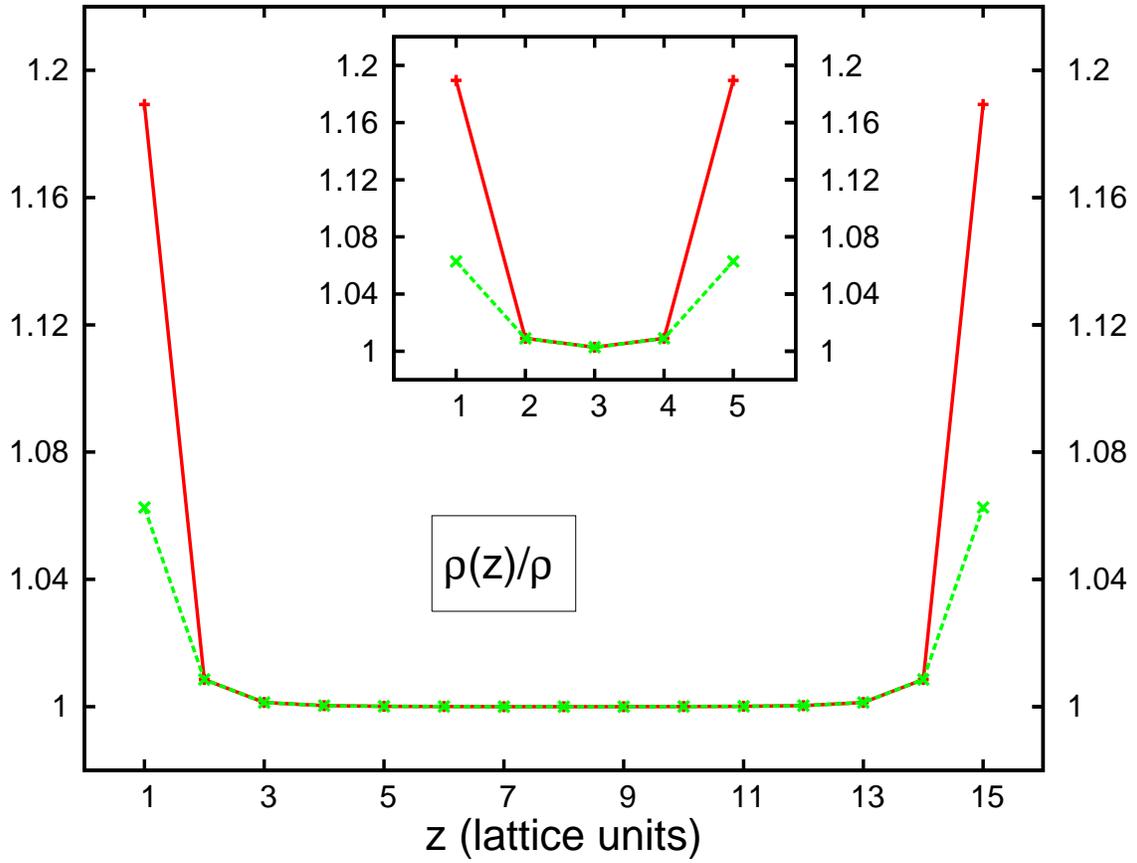,width=1.0\linewidth,height=0.5\textheight}
		\caption{\label{fig_densityprofile} (Color online) Transverse density profile for water between two hydrophobic surfaces separated by distance $L = 16$ and (inset) $L = 6$. Here, $h = 3.58$. The steeper (red) curve corresponds to ideal case $\nu_{S_{1}} = \nu_{S_{2}} = 0$ and the other (green) curve corresponds to $\nu_{S_{1}} = \nu_{S_{2}} = -0.5$.}
	\end{figure*}

The transverse density profile is shown in Fig.(\ref{fig_densityprofile}) after scaling  $\rho(z)$ with respect to bulk density value. At both interfaces there is a characterstic rise in density. From expressions of $\rho_{C}(z)$ (Appendix \ref{app_densityprofile}) and interface terms (Eq.\ref{average_Gamma}) it is evident that net contribution of $\phi$ field correlations is numerically small since density is charge-neutral quantity and linear $\phi$-dependent terms tend to cancel each other. Hence, away from interfaces density reaches bulk density value rapidly within a distance $\xi_{\eta}$. Many a model simulations in the past computed the transverse density profile for water confined between model hydrophobic surfaces. The short distance density increase is generically observed \citep{Paschek2001,*PradeepPRE2005,GardePNAS2009}. At ambient conditions the magnitude of interfacial density is seen to be typically $1.3$ times bulk density value near surfaces with alkane headgroups \citep{GardePNAS2009} and independent of $L$. In our model study we see an $L$-independent increase of magnitude $1.2$ for an ideal hydrophobic surface. The under-estimation could possibly be due to discrete orientational freedom envisaged in our model. Also, alkane head-groups in simulations may have an extra entropy due to fluctuating short length polymer chains.

The rise in interfacial density is also seen for water in the vicinity of hydrophilic surfaces \citep{GardePNAS2009,BerkowitzJCP2006}. In our model study $\rho(z)$ between hydrophilic surfaces also displays qualitatively similar profile and a lower magnitude of interfacial density compared to that near an ideal hydrophobic surface. In all cases the phenomenon is seen to be a consequence of the fact that water density has to vanish on the surface. This is compensated by an increase at the interface and the system comes back to its bulk equilibrium density within a distance $\xi_{\eta}$ from the interface.

We also calculate density correlations within the interfacial plane and between sites on interface and away from interface. Density correlations between any two sites $r$ and $r^{'}$ can be calculated from :
	\begin{equation}
		\langle W(r) W(r^{'}) \rangle \ = \ \left\langle \frac{\mu \lbrace \ldots \rbrace}{Z_{site}(r)} \frac{\mu \lbrace \ldots \rbrace}{Z_{site}(r^{'})} \right\rangle
	\end{equation}
where, $Z_{site}(r)$ is site functional at $r$. To compute density correlations on same interface, the site functional at both sites is given by Eq.(\ref{interfaceaction}). For density correlations between a site on interface and another, away from interface, the site functionals are given by Eqs.(\ref{interfaceaction}, \ref{zsite}) respectively. $\mu \lbrace \ldots \rbrace$ refers to the term proportional to $\mu$ in the respective site functional. The connected part of the correlation is given by $<W(r) W(r^{'})>_{c} \ \equiv ( <W(r) W(r^{'})> - <W(r)> <W(r^{'})> )$. The explicit expression in each context is deduced upto one-loop order in terms of ${\cal G}_{\eta}$, ${\cal G}_{\phi}$ and are given in Appendix (\ref{app_interface_bulk_correlations}).

\begin{figure*}
		\epsfig{file=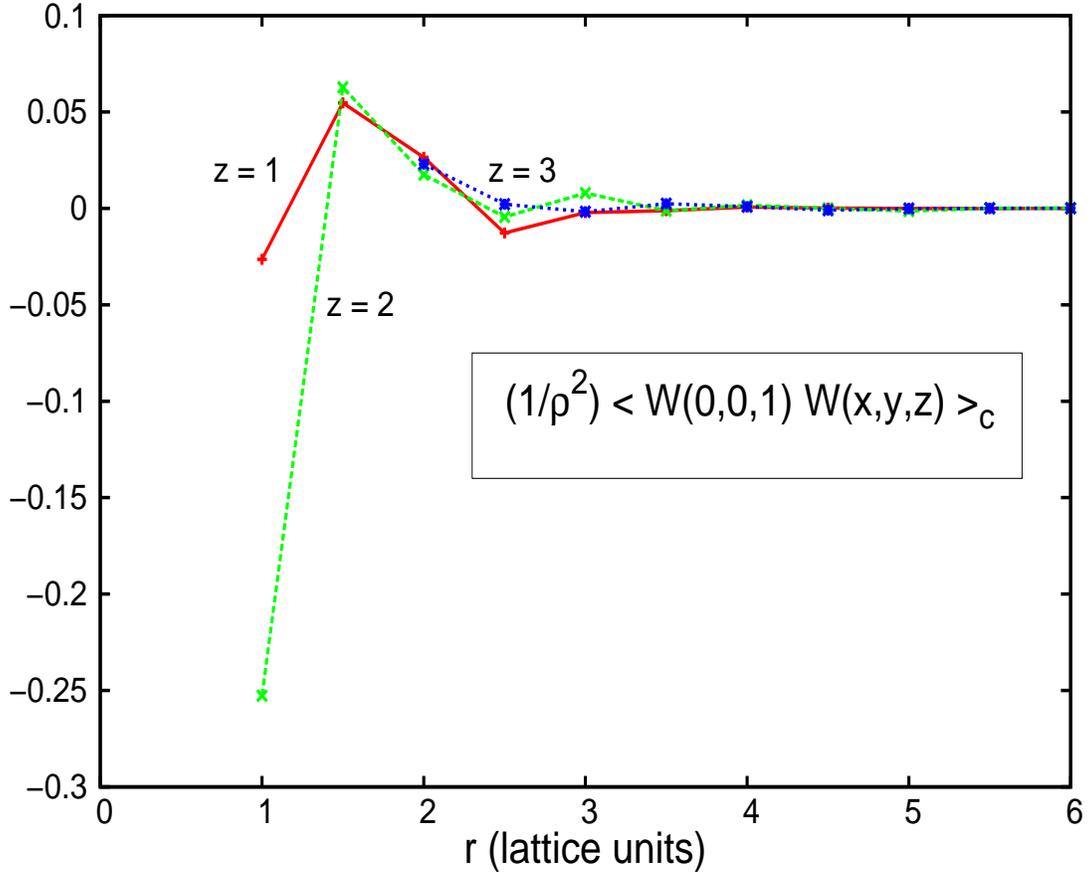,width=0.9\linewidth,height=0.5\textheight}
		\caption{\label{fig_ww} (Color online) Density correlations near a hydrophobic interface ($\nu_{S} = -0.5$) scaled appropriately with respect to bulk density value at $h = 3.58$. Correlations are between a reference site on interface ($z = 1$) and an arbitrary site on a plane defined by its $z$ coordinate. Distance between the two sites is measured using Euclidean metric. }
	\end{figure*}

Density correlations scaled appropriately with respect to bulk density value are plotted in Fig.(\ref{fig_ww}). The plot corresponds to $h = 3.58$. The figure essentially indicates density correlations do not extend beyond few molecular diameters from the interface. Also, there is no significant difference between correlations within an interface and that of between interface and non-interface sites. 

Similarly, orientational correlations can also be analyzed using the expressions for orientational weights given in Appendix (\ref{app_orientationalweight}). Their effect persists upto longer distance away from interface, proportional to the long correlation length of $\phi$ field.

\appendix

\section{Correlation functions in bulk water}\label{app_correlationfunctions}

The precise expressions for propagator functions $P_{\eta \eta}$, $P_{\phi \phi}$ deduced in MMF theory are :
	\begin{align}
		\nonumber P_{\eta \eta}(\Delta) \  = \ \left[ 64 \mu^{'}  \Delta^{2} \left( \frac{9}{10} - \mu^{'} \right) + 64 \mu^{'}\Delta  \left( - \frac{9}{10} + \mu^{'} - \frac{\nu^{'}}{4} - \frac{\lambda^{'}}{2} \right)  \right.
		\\  \displaystyle \left. + { \nu^{'}} + 4 \lambda^{'} + 16 \mu^{'} -  \left( \nu^{'} + 2 \lambda^{'} - 4 \mu^{'}  \right)^{2} \vphantom{\frac{9}{10}} \right] &
	\end{align}
	\begin{equation}	
		P_{\phi \phi}(\Delta) \ = \ \left[ \frac{\displaystyle 96 \mu^{'}}{\displaystyle 5}   \Delta \left( 1 - \Delta \right) + {\displaystyle \nu^{'}} \right]
	\end{equation}
where, $\Delta = \displaystyle \frac{1}{6} \sum_{i=1}^{3} (1 - \text{cos}(k_{i}))$ and $\nu^{'} = 2 \nu/Z_{o}$, $\lambda^{'} = \lambda/Z_{o}$, $\mu^{'} = 90 \mu/Z_{o}$ are scaled fugacities. They vary between $0$ and $1$. From zeroth order partition function, $\textnormal{DB} = \nu^{'}$, $\textnormal{HB} = \lambda^{'}$, $\rho = \mu^{'}$.

The Green's functions for $\eta$, $\phi$ fields in bulk water are given by :
	\begin{equation}
		{\cal G}_{\eta}(\vec{r_{1}}, \vec{r_{2}}) \ = \ \int\limits_{-\pi}^{\pi} \frac{ d\vec{k} }{(2\pi)^{3}} \frac{ \exp\left( i \vec{k} \cdot (\vec{r_{1}} - \vec{r_{2}}) \right) }{ P_{\eta \eta}(\vec{k}) }
	\end{equation}
and similarly for $\phi$ field. $\vec{r_{1}}$, $\vec{r_{2}}$ are position indices for any two sites.

For large $r = | \vec{r_{1}} - \vec{r_{2}} |$, ${\cal G}_{\eta}(r)$ is of the functional form,
	\begin{equation}
		{\cal G}_{\eta}(r) \ \propto \ \frac{\exp\left( -{r}/{\xi_{\eta}} \right) }{r} \sin(\omega_{\eta} r)
	\end{equation}
where, to the leading order, 
	\begin{align}
		(\xi_{\eta})^{-1} \ & = \ \displaystyle \sqrt[4]{\frac{ \frac{3}{8}}{\frac{9}{10} - \rho } } \ \sin\left(\frac{1}{2}{\tan}^{-1} \sqrt{\frac{50}{27}\left(\frac{9}{25} - \rho \right) } \ \right)
		\\ \omega_{\eta} \ & = \ \displaystyle \sqrt[4]{ \frac{ \frac{3}{8}}{\frac{9}{10} - \rho } } \ \cos\left(\frac{1}{2}{\tan}^{-1} \sqrt{\frac{50}{27}\left(\frac{9}{25} - \rho \right) } \ \right)
	\end{align}	
The form of ${\cal G}_{\eta}$ implies periodic peaks whose amplitudes fall-off exponentially with distance. The ${\cal G}_{\phi}$ correlator, in addition to oscillatory behavior at short distances, takes the following asymptotic form for large $r$ :
	\begin{equation}
		 {\cal G}_{\phi}(r) \ \propto \ \frac{\exp\left(- {r}/{\xi_{\phi}} \right)}{r}
	\end{equation}
	where,
	\begin{align}	 
		(\xi_{\phi})^{-1} \  = \ \sqrt{6 \left( \sqrt{1+ \frac{5 \ \textnormal{DB}}{24 \rho}} - 1 \right)} 
		\ = \ \sqrt{6 \left( \sqrt{1+ \frac{5}{24}(4 - h)} - 1 \right)} \label{phicorrelation}
	\end{align}
The above expression for $\xi_{\phi}$ is given to the leading order.

\section{Orientational weight $C(\eta, \phi)$ } \label{app_orientationalweight}

The orientational weight for water state in bulk water is given by :

	\begin{equation}		
		\nonumber C(\eta, \phi) \ =  \sum^{'}_{\displaystyle \stackrel{\displaystyle H_{\alpha} = 0,\pm 1}{\alpha = \pm 1, \pm 2, \pm 3}} \exp\left[ i \sum_{\alpha} ( H^{2}_{\alpha}(r) {\eta}(r+e_{\alpha}) + H_{\alpha}(r) {\phi}(r+e_{\alpha}) )  \right]
	\end{equation}
where, prime indicates summation is subject to constraints Eqs.(\ref{H2constraint},\ref{Hconstraint}).

About the mean field configuration $\eta = \phi = 0$ fields are expanded upto quadratic order. The $C(\eta, \phi)$ is then given by :
	\begin{equation}
		C(\eta, \phi)  \simeq \ 90 \left[ 1 + \frac{2 i}{3} \sum_{\alpha} \eta_{\alpha} - \frac{1}{3} \sum_{\alpha} \eta_{\alpha}^{2}   - \frac{2}{5} \sum_{\alpha, \beta} \eta_{\alpha}  \eta_{\beta} - \frac{1}{3} \sum_{\alpha} \phi_{\alpha}^{2} + \frac{2}{15} \sum_{\alpha, \beta} \phi_{\alpha}  \phi_{\beta} \right]
	\end{equation}
where $\eta_{\alpha} \equiv \eta(r + e_{\alpha})$ and $\phi_{\alpha} \equiv \phi(r + e_{\alpha})$ . 

The orientational weights for affected orientations of an interfacial water near hydrophobic surface are denoted by $C^{'}(\eta, \phi)$. With boundary condition $\eta = \phi = 0$ on surface sites, it is given by :
	\begin{equation}
		C^{'}(\eta, \phi) \simeq \ 60 \left[ 1 + \frac{3 i}{5} \sum_{\alpha} \eta_{\alpha} - \frac{3}{10} \sum_{\alpha} \eta_{\alpha}^{2}   - \frac{3}{10} \sum_{\alpha, \beta} \eta_{\alpha}  \eta_{\beta} - \frac{3}{10} \sum_{\alpha} \phi_{\alpha}^{2} + \frac{1}{10} \sum_{\alpha, \beta} \phi_{\alpha}  \phi_{\beta} \right]
	\end{equation}
Near a hydrophilic surface, 
	\begin{align}
		\nonumber C^{'}_{\pm}(\eta, \phi) \ \simeq \ 30 \left[ 1 + \frac{3 i}{5} \sum_{\alpha} \eta_{\alpha} - \frac{3}{10} \sum_{\alpha} \eta_{\alpha}^{2}   - \frac{3}{10} \sum_{\alpha, \beta} \eta_{\alpha}  \eta_{\beta} - \frac{3}{10} \sum_{\alpha} \phi_{\alpha}^{2} \right. &
		\\ \left. + \frac{1}{10} \sum_{\alpha, \beta} \phi_{\alpha}  \phi_{\beta} \mp \frac{i}{5} \sum_{\alpha} \phi_{\alpha}    \right] &
	\end{align}

\section{Density profile} \label{app_densityprofile}

$\rho_{C}(z) $ in Eq.(\ref{rhoz}) is given by,
	\begin{equation}
		\rho_{C}(z) \ \equiv \ \frac{ \partial (\ln (Z)) }{ \partial (\beta \tilde{\mu}(z)) } \ = \ \mu^{'} - \frac{1}{2} \left[ (-\nu^{'} \mu^{'}) T_{1} + (- \lambda^{'} \mu^{'}) T_{2} + \mu^{'} (1 - \mu^{'}) T_{3} \right]
	\end{equation}
where,
	\begin{subequations}	
	\begin{align}
		& T_{1} \ = \ (1 - 2 (\nu^{'} + 2 \lambda^{'})) {\cal G}_{\eta}(r,r) + \frac{ 4 \mu^{'} }{3} \sum_{\alpha}{\cal G}_{\eta}(r,r + e_{\alpha})  + {\cal G}_{\phi}(r,r)
		\\ & T_{2} \ = \ (4 - 4(\nu^{'} + 2 \lambda^{'})) {\cal G}_{\eta}(r,r) + \frac{ 8 \mu^{'} }{3} \sum_{\alpha}{\cal G}_{\eta}(r,r + e_{\alpha})
		\\ & \nonumber T_{3} \ = \ \frac{ 4 \mu^{'} }{3} (\nu^{'} + 2 \lambda^{'}) \sum_{\alpha}{\cal G}_{\eta}(r,r + e_{\alpha}) + \frac{4}{15} \sum_{\alpha}{\cal G}_{\eta}(r + e_{\alpha},r + e_{\alpha}) + \frac{4}{5} \sum_{\alpha}{\cal G}_{\phi}(r + e_{\alpha},r + e_{\alpha})
		\\ & \qquad  + \frac{2}{15} \left( 1 - \frac{20 \mu^{'}}{9} \right) \sum_{\alpha,\alpha{'}}{\cal G}_{\eta}(r + e_{\alpha},r + e_{\alpha^{'}}) - \frac{2}{15} \sum_{\alpha,\alpha^{'}}{\cal G}_{\phi}(r + e_{\alpha},r + e_{\alpha^{'}})
	\end{align}
	\end{subequations}
where, $r = (x,y,z)$ is a site position, $r + e_{\alpha}$ is a near-neighbor site in $e_{\alpha}$ direction. The Green's functions ${\cal G}_{\eta}(r_{1},r_{2})$, ${\cal G}_{\phi}(r_{1},r_{2})$ are computed using Eq.(\ref{Greensfunction}). The scaled fugacities $\nu^{'}$, $\lambda^{'}$, $\mu^{'}$ are as defined in Appendix (\ref{app_correlationfunctions}).

\section{Density correlations} \label{app_interface_bulk_correlations}

The connected part of density correlation between sites on same interface, to the leading order, is given by the expression : 
	\begin{align}
		\nonumber \langle W(r) W(r^{'}) \rangle_{c} \  \simeq \ 
		 -  (\mu^{'})^{2} & \left[ (\nu^{'} + 2 \lambda^{'} )^{2} {\cal G}_{\eta}(r,r^{'})  + \left( \frac{2}{3} \left( 1 + \frac{3 \nu_{S}}{5}  - \mu^{'} \right) \right)^{2} \sum_{\alpha,\alpha^{'}}{\cal G}_{\eta}(r + e_{\alpha},r^{'} + e_{\alpha^{'}}) \right.
		\\ &  \left. + \ 2 \left( \frac{2}{3} \left( 1 + \frac{3  \nu_{S}}{5} - \mu^{'} \right) \right)(\nu^{'} + 2 \lambda^{'} ) \sum_{\alpha^{'}}{\cal G}_{\eta}(r,r^{'} + e_{\alpha^{'}})   \right] \label{ww_interface}
	\end{align}
where, $r$, $r^{'}$ are arbitrary sites on same interface; $r + e_{\alpha}$, $r^{'} + e_{\alpha^{'}}$ are respective near-neighbor sites in the directions $e_{\alpha}$, $e_{\alpha^{'}}$ respectively. ${\cal G}_{\eta}(r_{1},r_{2})$, ${\cal G}_{\phi}(r_{1},r_{2})$ can be computed from Eq.(\ref{Greensfunction}). The scaled fugacities $\nu^{'}$, $\lambda^{'}$, $\mu^{'}$ are as defined in Appendix (\ref{app_correlationfunctions}).

The density correlation between a site on interface and another, away from interface is given by,
	\begin{align}
		\nonumber & \langle W(r) W(r^{'}) \rangle_{c} \quad \simeq \ 
		\\ \nonumber & - (\mu^{'})^{2} \left[ (\nu^{'} + 2 \lambda^{'} )^{2} {\cal G}_{\eta}(r,r^{'})  +  \frac{2}{3}(1 - \mu^{'}) \left( 1 + \frac{3 \nu_{S}}{5}  - \mu^{'} \right) \sum_{\alpha,\alpha^{'}}{\cal G}_{\eta}(r + e_{\alpha},r^{'} + e_{\alpha^{'}}) \right.
		\\ & \left. \qquad  + \ \frac{2}{3} (\nu^{'} + 2 \lambda^{'} ) \left( 2 + \frac{3  \nu_{S}}{5} - 2 \mu^{'} \right)  \left( \sum_{\alpha} {\cal G}_{\eta}(r + e_{\alpha},r^{'}) +  \sum_{\alpha^{'}}{\cal G}_{\eta}(r,r^{'} + e_{\alpha^{'}}) \right)  \right] 
		\label{ww_interface_bulk}
	\end{align}
where, $r$ is any site on interface, $r^{'}$ is away from interface; $r + e_{\alpha}$, $r^{'} + e_{\alpha^{'}}$ are their respective near-neighbor sites. 

Orientational correlations can also be evaluated using the expression for orientational weights given in Appendix (\ref{app_orientationalweight}) and can be computed using known expressions for Green's functions.

\begin{acknowledgments}

We thank Dr. Gautam Menon for bringing transverse density profile to our attention.

\end{acknowledgments}

	\begin{figure*}[h]
		\epsfig{file= ,width=0.75\linewidth,height=0.4\textheight}
		\caption{\label{fig_allowedconfigs} Allowed configurations : a water site with two hydrogen arms (+) and two lone-pair arms (-) on links around the site (Eqs.\ref{H2constraint},\ref{Hconstraint}). A hydrogen bond occurs when a hydrogen arm (+) and a lone-pair arm (-) of two molecules meet at a site. (right bottom corner)  unit vectors on cubic lattice.}
	\end{figure*}

\end{document}